\documentclass[useAMS,usenatbib]{mn2e}

\usepackage{color}
\usepackage{amssymb}
\usepackage[pdftex]{graphicx}
\usepackage{epstopdf}
\usepackage[draft]{hyperref}
\hypersetup{colorlinks,breaklinks,linktoc=page,linkcolor=black,citecolor=blue   } 
\usepackage[T1]{fontenc}  
\usepackage{aecompl}
\usepackage{soul}

\usepackage{eso-pic}

\AddToShipoutPictureBG*{%
  \AtPageUpperLeft{%
    \hspace{17cm}%
    \raisebox{-3.5\baselineskip}{%
      \makebox[0pt][l]{\textnormal{DES 2015-0083}}
}}}%

\AddToShipoutPictureBG*{%
  \AtPageUpperLeft{%
    \hspace{15cm}%
    \raisebox{-4.5\baselineskip}{%
      \makebox[0pt][l]{\textnormal{FERMILAB-PUB-16-002-AE}}
}}}%

\newcommand{\reff}[1]{eq. ({\ref{#1}})}

\newcommand{\Msun}{\mbox{$M_{\odot}$}}

\newcommand{\lephare}{\textsc{LePhare}$\;$}

\title[Observations of RXJ2248 in DES and CLASH]{Comparing Dark Energy Survey and \emph{HST}--CLASH observations of the galaxy cluster RXC J2248.7--4431: implications for stellar mass versus dark matter }
\author[A. Palmese et al]{\parbox{\textwidth}{ 
A. Palmese$^{1}$\thanks{E-mail: apalmese@star.ucl.uk},  O. Lahav$^{1}$,  M. Banerji$^{2}$, D. Gruen$^{{3},{4},{5},{6}}$, S. Jouvel$^{1}$,  P. Melchior$^{7}$, J. Aleksi\'c$^{8}$, J. Annis$^{9}$, H. T. Diehl$^{9}$, W. G. Hartley$^{1}$, T. Jeltema$^{10}$, 
A. K. Romer$^{11}$, 
E. Rozo$^{12}$, 
E. S. Rykoff$^{{3},{4}}$, 
S. Seitz$^{{5},{6}}$, 
E. Suchyta$^{13}$, 
Y. Zhang$^{14}$, 
T. M. C.~Abbott$^{15}$,
F.~B.~Abdalla$^{{1},{16}}$,
S.~Allam$^{9}$,
A.~Benoit-L\'evy$^{{17},{1},{18}}$,
E.~Bertin$^{{17},{18}}$,
D.~Brooks$^{1}$,
E.~Buckley-Geer$^{9}$,
D.~L.~Burke$^{{4},{3}}$,
D.~Capozzi$^{19}$,
A.~Carnero~Rosell$^{{20},{21}}$,
M.~Carrasco~Kind$^{{22},{23}}$,
J.~Carretero$^{{24},{8}}$,
M.~Crocce$^{24}$,
C.~E.~Cunha$^{4}$,
C.~B.~D'Andrea$^{{19},{25}}$,
L.~N.~da Costa$^{{20},{21}}$,
S.~Desai$^{{26},{27}}$,
J.~P.~Dietrich$^{{26},{27}}$,
P.~Doel$^{1}$,
J.~Estrada$^{9}$,
A.~E.~Evrard$^{{28},{14}}$,
B.~Flaugher$^{9}$,
J.~Frieman$^{{9},{29}}$,
D.~W.~Gerdes$^{14}$,
D.~A.~Goldstein$^{{30},{31}}$,
R.~A.~Gruendl$^{{22},{23}}$,
G.~Gutierrez$^{9}$,
K.~Honscheid$^{{32},{33}}$,
D.~J.~James$^{15}$,
K.~Kuehn$^{34}$,
N.~Kuropatkin$^{9}$,
T.~S.~Li$^{35}$,
M.~Lima$^{{36},{20}}$,
M.~A.~G.~Maia$^{{20},{21}}$,
J.~L.~Marshall$^{35}$,
C.~J.~Miller$^{{28},{14}}$,
R.~Miquel$^{{37},{8}}$,
B.~Nord$^{9}$,
R.~Ogando$^{{20},{21}}$,
A.~A.~Plazas$^{38}$,
A.~Roodman$^{{3},{4}}$,
E.~Sanchez$^{39}$,
V.~Scarpine$^{9}$,
I.~Sevilla-Noarbe$^{{39},{22}}$,
R.~C.~Smith$^{15}$,
M.~Soares-Santos$^{9}$,
F.~Sobreira$^{{9},{20}}$,
M.~E.~C.~Swanson$^{23}$,
G.~Tarle$^{14}$,
D.~Thomas$^{19}$,
D.~Tucker$^{9}$,
V.~Vikram$^{40}$
\vspace{0.3cm} \\
\emph{\small(Affiliations are listed at the end of paper)} 
}}

\begin{document}

\pagerange{\pageref{firstpage}--\pageref{lastpage}} \pubyear{2016}

\maketitle

\label{firstpage}

\begin{abstract}
We derive the stellar mass fraction in the galaxy cluster RXC J2248.7--4431 observed with the Dark Energy Survey (DES) during the Science Verification period.  We compare the stellar mass results from DES (five filters) with those from the \emph{Hubble Space Telescope} Cluster Lensing
And Supernova Survey (CLASH; 17 filters). When the cluster spectroscopic redshift is assumed, we show that stellar masses from DES can be estimated within 25\% of CLASH values. 
We compute the stellar mass contribution coming from red and blue galaxies, and study the relation between stellar mass and the underlying dark matter using weak lensing studies with DES and CLASH.  An analysis of the radial profiles of the DES total and stellar mass yields a stellar-to-total fraction of $f_\star=(6.8\pm 1.7)\times10^{-3}$ within a radius of $r_{200c}\simeq 2$ Mpc. Our analysis also includes a comparison of photometric redshifts and star/galaxy separation efficiency for both data sets. 
We conclude that space-based small field imaging can be used to calibrate the galaxy properties in DES  for the much wider field of view. The technique developed to derive the stellar mass fraction in galaxy clusters can be applied to the $\sim100\, 000$ clusters that will be observed within this survey and yield important information about galaxy evolution.
\end{abstract}


\section{Introduction}
In the last decade, large photometric galaxy surveys, such as Sloan Digital Sky Survey (SDSS), have provided us with a massive amount of data that have proven to be extremely useful for studies of cosmology. On the other hand, smaller area but deeper surveys like the \emph{Hubble Space Telescope} (HST) based Cluster Lensing And Supernova Survey (CLASH) (\citealt{postman}) allowed us to characterize single objects with unprecedented precision. The importance of finding synergies between these surveys relates to several aspects of observation (e.g. target selection, photometric calibration) and data analysis (photometric redshifts, physical properties of galaxies). This is particularly relevant for overlapping ground-- and space--based surveys: the higher quality that can be obtained from space can enable calibration and tests for the data collected by ground--based telescopes. 

In this paper, we study the cluster of galaxies RXC J2248.7--4431 (hereafter RXJ2248). We make use of the synergies between DES and CLASH, and test in this way the performance of the early DES data at a catalogue level (i.e. without making use of the images for the results). Photometric redshift (photo-$z$) and stellar mass results from CLASH are also used as a validation set for DES stellar mass estimates.

The aim of this paper is twofold: the first goal is to compare between DES's wide area breadth and CLASH's small area precision for the cluster RXJ2248. In fact, checks using \emph{HST} data had not been done before to test the DES data, although the similar optical filters and the additional UV and IR \emph{HST} bands make CLASH an optimal candidate for validation  and quantifying uncertainties of photometry, photo-$z$'s and stellar masses. The second is to illustrate how an analysis of the stellar mass distribution of this massive cluster over the wider Dark Energy Camera (DECam) field of view can be done. 

In Section \ref{observations}, we start by describing the two surveys considered. The cluster is described in Section \ref{previousworks}. The comparison of DES and CLASH, in terms of photometric redshifts and star/galaxy separation, is presented in Section \ref{comparison}. Section \ref{DMandSM} contains the second part of this paper, where we present the stellar mass results obtained from CLASH and DES, and compare the DES stellar masses to the total mass from the DES weak lensing analysis by \citet{melchior}. 
In the following, we assume a concordance $\Lambda$CDM cosmological model with $\Omega_m =0.3$, $\Omega_\Lambda =0.7$ and $h=0.7$. In this cosmology, 1 arcmin corresponds to a physical transverse length of 295 kpc at the cluster redshift $z=0.3475$. 

The notation adopted in this paper for the cluster mass and radius follows the one often used in literature. The radii of spheres around the cluster centre are written as $r_{\Delta m}$ and $r_{\Delta c}$ where $\Delta$ is the overdensity of the sphere with respect to the mean matter density (subscript $m$) or the critical density (subscript $c$) at the cluster redshift. Masses inside those spheres are therefore $M_{\Delta m}=\Delta \frac{4\upi}{3} r^3_{\Delta m}\rho_m$ and similarly for $M_{\Delta c}$.  In the following, we quote $\Delta=200$, which is the density contrast at virialization for a dark matter halo.

\section{Data}\label{observations}
The data used for the analyses developed for this paper come from DES and CLASH. The Dark Energy Survey (DES) is an optical-near-infrared survey that is imaging 5000 ${\rm deg}^2$ of
the South Galactic Cap in the $grizY$ bands over 525 nights spanning 5 years. The survey is being carried out using a new $\sim 3$ $\textrm{deg}^2$ CCD
camera (the DECam, see \citealt{flaugher}) mounted on the Blanco 4-m telescope at the Cerro Tololo Inter-American Observatory (CTIO) in Chile. DES started in 2012 with a testing period (November 2012 -- February 2013) called DES Science Verification (SV)\footnote{For public data release see: \url{http://des.ncsa.illinois.edu/releases/sva1}}. At the time of writing, two observing seasons (\citealt{y1}) have been completed, and a third is underway.

The survey strategy is designed to optimize the photometric calibration by tiling each region of the survey with several overlapping pointings in each band. This provides uniformity of coverage and control of systematic photometric errors. This strategy allows DES to determine photometric redshifts of galaxies to an accuracy of $\sigma(z) \simeq 0.07 $ out to $z \gtrsim 1$, with some dependence on redshift and galaxy type, and cluster photometric redshifts to $\sigma(z) \sim 0.02$ or better out to $z \simeq 1.3$ (\citealt{descollaboration}). It will also provide shapes for approximately 200 million galaxies for weak lensing studies. For further information, see \citet{descollaboration} or \url{www.darkenergysurvey.org}. 

The fact that DECam has a $\sim 3$ $\textrm{deg}^2$ field of view gives us the opportunity of studying the large scale structure of galaxy clusters with only one pointing. 

\begin{table}\centering
\begin{tabular}{cccc}
$\alpha_{J2000}$&$\delta_{J2000}$&Redshift& Luminosity (erg $\mathrm{s}^{-1}$)\\
\hline
22:48:44.29 &-44:31:48.4 & 0.348&$ 3.08 \times 10^{45}$
\\
\end{tabular}\caption{Main properties of the cluster RXC J2248-4431. The quoted luminosity is in the rest frame 0.1-2.4 keV band.}\label{properties}
\end{table}
\begin{table*}
{\small
\centering
\begin{tabular}{c|cccccccc}
Survey/Instrument& Authors & FoV &Filters& Mag limits&Spectra&Objects\\
\hline
NTT+GMOS&\citet{gomez}&$5'\times 5'$&$V$, $R$&--&116&711\\
CLASH&\citet{postman},&$3.4'\times3.4'$ (ACS), &16 in $2000-17000\, \mathrm{\AA}$&$\sim$ 25-27 ($10\sigma$)&-- &3471\\
& \citet{monna} &$2'\times2'$ (WFC3)&&&&\\
WFI&\citet{gruen} &$33'\times 33'$& $UBVRIZ$&26.4, 26.7 , 24.4 &--&--\\
&&&&($VRI$ $5\sigma$)&&&\\
DES SV& \citet{melchior}&2.2\,$\mathrm{deg}^2$&$grizY$& 24.45, 24.30, 23.50,&--&$374\;294$\\
&&&&22.90, 21.70 (10$\sigma$)&&&\\

\end{tabular}}\caption{Some experimental specifications of the surveys that have observed RXJ2248, with the corresponding paper in which those data have been used. The work presented in those papers is briefly summarized in Section \ref{previousworks}. The magnitude limits reported for DES are the mean 10$\sigma$ galaxy magnitudes.}\label{surveys}
\end{table*}

The cluster RXJ2248 was observed during the SV season, with typical exposure times of 90 s for the $griz$ bands and 45 s for the $Y$ band. It was re-observed later in 2013 to benefit from improvements to telescope performance and general image quality. The data reduction was done using the SVA1 DES Data Management (DESDM) pipeline, described in detail in \citet{sevilla}, \citet{desai} and \citet{dataproc}. The process includes calibration of the single-epoch images, that are then co--added after a background subtraction and cut into tiles. The SVA1 catalogue was created using \textsc{Source Extractor (SExtractor}, \citealt{sextractor}) to detect objects on the $riz$ co-added images. 
The median $10\sigma$ depths of SV data are $g\sim24.45, r\sim24.30,i\sim23.50,z\sim22.90, Y\sim21.70$, which reach close to the expected DES full depths. Limiting magnitudes were estimated for the 200 deg$^2$ SPT-E part of the wide--field SV area using \textsc{Balrog} (\citealt{suchyta}) and PSF magnitude errors for true point sources. 

We use AB magnitudes throughout this paper, and \texttt{MAG\_AUTO} measurements given by \textsc{SExtractor}, as these proved to be robust and were thus used in several DES SV papers (e.g. \citealt{bonnett},\citealt{crocce}). The objects selected for the analysis have a signal-to-noise ratio $(S/N)>10$ in the $i$ band.

The other survey considered here is CLASH (\citealt{postman}), a 524-orbit \emph{HST} multi-cycle treasury programme that has observed 25 massive clusters, having a range of virial masses between $5 \times 10^{14} M_\odot$ to $30 \times 10^{14} M_\odot$ and an average redshift of $\bar{z}=0.4$. The wavelength range covers the UV, the visible and the IR ($2000 - 17000 $ \AA) through 17 bands using the Advanced Camera for Surveys (ACS) and the Wide Field Camera 3 (WFC 3).

The CLASH mosaics were produced using the ``MosaicDrizzle'' pipeline (see \citealt{koekemoer}). The CLASH catalogue creation pipeline makes use of \textsc{SExtractor}: the software is run in dual image mode, where a detection image is created from a weighted sum of the ACS/WFC and WFC3/IR images.  The WFC3/UVIS images are not used in the construction of the detection image but the UVIS data are still used to compute source photometry. The photometry given in the public catalogue (\url{http://www.stsci.edu/~postman/CLASH}), which is also the one used in this work, was measured in isophotal apertures, as they have been shown to produce reliable colours (\citealt{benitez}). ACS/WFC3 reach a depth of 26.8, 26.4, 26.2, 26.0 and 26.6 ($10\sigma$ galaxy AB magnitudes for circular apertures of 0.4 arcsec in diameter, \citealt{postman}) in the F475W, F625W, F775W, F850LP and F105W filters, respectively.

Below, we compare the information obtained with five DES filters and with 17 \emph{HST} filters.

\section{The Cluster RXC J2248.7--4431}\label{previousworks}
In this section we present what is known about this cluster from previous works. The cluster of galaxies RXC J2248.7--4431, where RXC stands for \emph{ROSAT} X-ray Cluster, is also known as Abell S1063 or MACS 2248--4431. It is a very luminous cluster, having an X--ray bolometric luminosity of $(6.95 \pm 0.1)\times 10^{45}\; {\rm erg\, s^{-1}}$ in the energy range $0.1-100$ keV (\citealt{maughan08}). Its properties are listed in Table \ref{properties}, while the experimental specifications of surveys that have observed RXJ2248 are quoted in Table \ref{surveys}. It was first catalogued by \citet{abell}, who counted 74 galaxies. Thanks to the \emph{ROSAT}-ESO Flux Limited X-ray (REFLEX) Galaxy Cluster survey, \citet{bohr} measured a spectroscopic redshift $z=0.3475$, which has been adopted in the recent literature and has also been confirmed in \citet{gomez}, who quoted a mean redshift of $z=0.3461^{+0.0010}_{-0.0011}$ for 81 members.

\citet{gomez} were the first to study in detail RXJ2248, even though it is the second most luminous cluster in the REFLEX survey (having a reported luminosity of $\sim 3.08 \times 10^{45}$ erg $\mathrm{s}^{-1}$ in the rest frame $0.1-2.4$ keV band). 

In \citet{gomez}, the cluster is presented as one of the hottest X-ray clusters known at that time. The high X-ray temperature, together with the high velocity dispersion, suggest a very massive cluster ($M_{200\mathrm{c}}>2.5 \times 10^{15} \Msun$) and/or a merger system. The merger model is supported by a small offset between the galaxy distribution and the peak of X-ray isophotes, and a non-Gaussian galaxy velocity distribution. \citet{gomez} also reported that the velocity distribution is better represented by the velocity dispersion produced during a merger than by the velocity distribution of a relaxed cluster.

\citet{gruen} used Wide-Field Imager (WFI) data to perform a weak lensing analysis of the cluster. They parametrized the cluster density with a NFW profile (\citealt{nfw}) and obtained a mass $M_{200m}=33.1^{+9.6}_{-6.8}\times 10^{14}\Msun$ (or $M_{200c} = 22.8 ^{+6.6}_{-4.7}\times 10^{14} \Msun$). They also identified a second galaxy cluster in the field of view at redshift $\sim 0.6$, with an estimated mass of $M_{200m}=4.0^{+3.7}_{-2.6}\times10^{14}\Msun$.

\citet{melchior} studied the weak lensing masses and galaxy distributions of four massive clusters observed during the DES SV period, including RXJ2248. They found $M_{200\mathrm{c}}=17.5^{+4.3}_{-3.7}\times 10^{14} \Msun$, which is in agreement with previous mass estimates. For RXJ2248, they also identified filamentary structures of the luminous red-sequence galaxies found with the \textsc{RedMaPPer} (\citealt{redpaper}) algorithm. 

\citet{umetsu15} combined \emph{HST} and wide field imaging (from the Subaru telescope or the ESO/WFI) observations to reconstruct the surface mass density profiles of 20 CLASH clusters. Their analysis jointly uses strong lensing as well as weak lensing with shear and magnification, and for RXJ2248 they found $M_{200c}=18.78 \pm 6.72 \times 10^{14}\Msun$.

\begin{figure}
\centering
\includegraphics[width=0.5\textwidth]{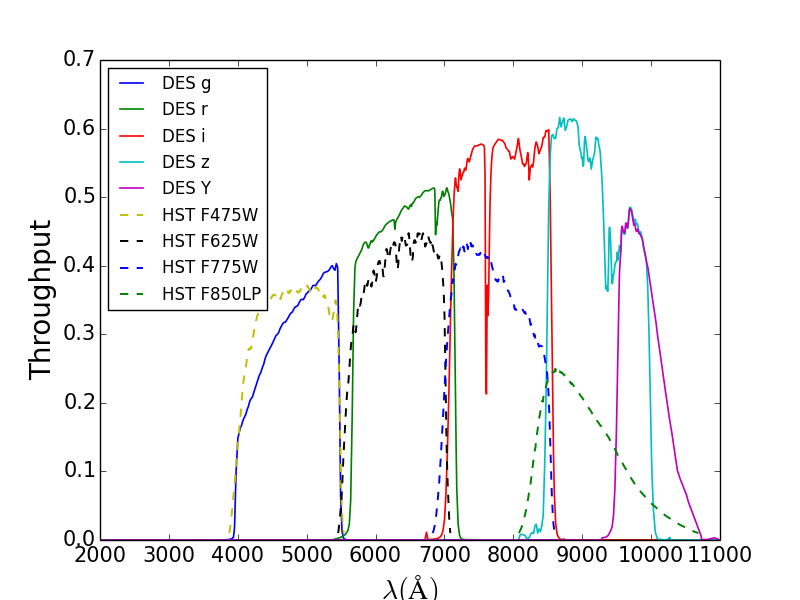}\caption{Throughput of the DES filters (solid lines) and \emph{HST} similar filters (dashed lines).}\label{filters}
\end{figure}
\begin{figure}\centering
\includegraphics[width=0.5\textwidth]{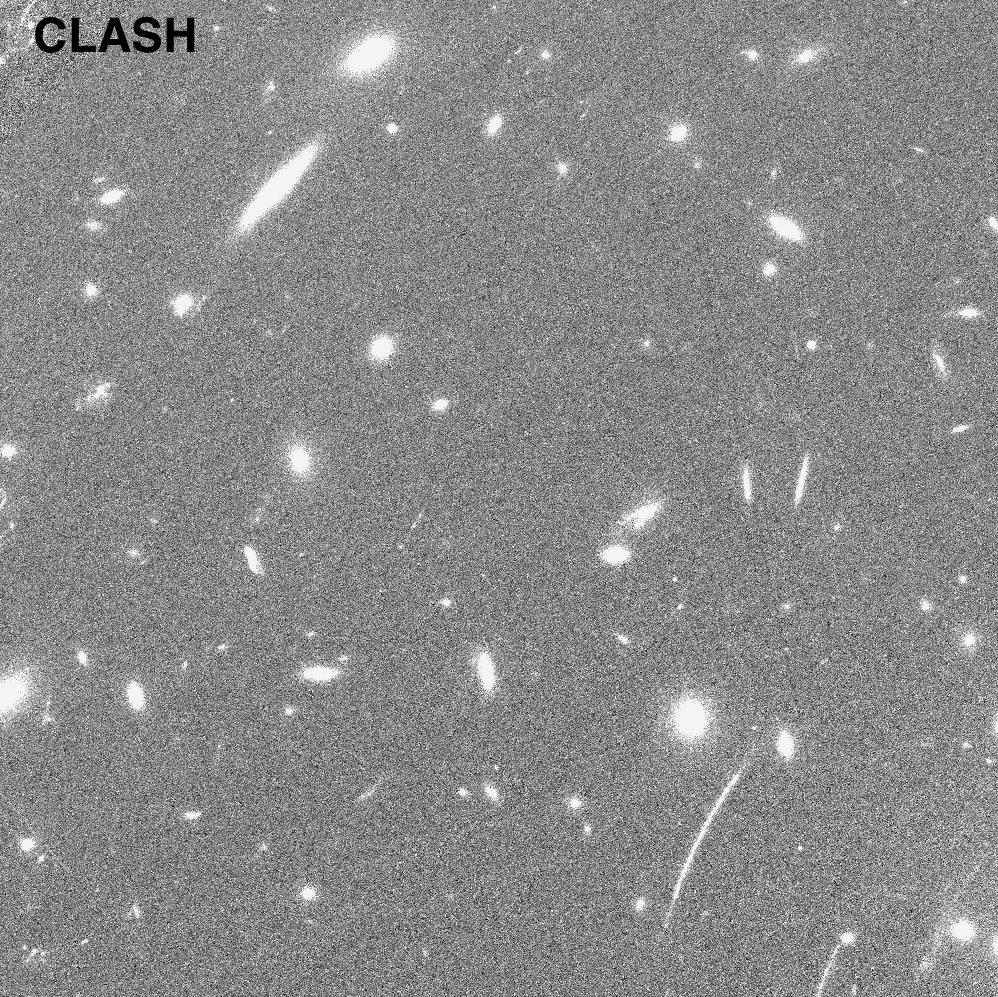}\\
\includegraphics[width=0.5\textwidth]{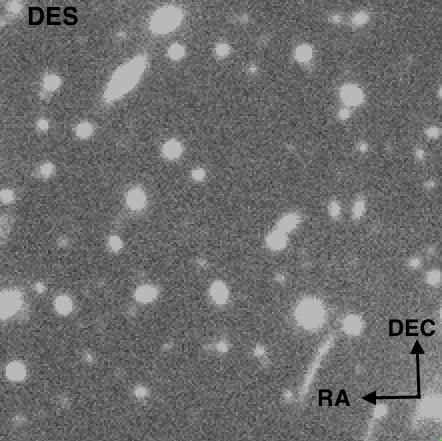}
\caption{A portion of $1'\times 1'$ image centred in RA 22:48:48.003 and DEC -44:31:38.52 in the CLASH F625W band (top) and the DES $r$ band (bottom).}\label{images}
\end{figure}

\section{Comparison of DES and CLASH}\label{comparison}
In this section, we assess detectability, photometry, and stellar masses of DES galaxies, treating matched CLASH galaxies as truth table.\footnote{A comparison of weak lensing measurements between DES and CLASH was not performed because they predominantly reveal differences in the shear calibration. The majority of galaxies with shape measurement in both catalogues are very faint for DES, resulting in large and noisy calibration factors (see Section 4.2.1 in \citealt{melchior}). In addition, the high density of galaxies in the central region of this cluster creates many more close galaxy pairs or even blends in ground--based DES images than when viewed with \emph{HST}, rendering shape measurement even more challenging. A detailed analysis of those relevant effects is beyond the scope of this paper.} In order to make the comparison, we seek to identify similar filters in both data sets. Figure \ref{filters} shows that the closest \emph{HST} analogs to DES $griz$ are F475W, F625W, F775W and F850lp. We will refer to the corresponding \emph{HST} and DES bands as $g$, $r$, $i$ and $z$ for simplicity of notation. In the following, we will also use the DES $Y$ band, which does not have a similar \emph{HST} filter. When we refer to 5 CLASH filters, it means we are including the F105W filter, that is broader than the DES $Y$. 

In the DES catalogue of the RXJ2248 area, there are $374\,294$ sources in a roughly circular area of approximately 3 $\mathrm{deg}^2$. 
The deeper, higher resolution CLASH catalogue includes $3\,471$ sources in a much smaller area ($\sim 5'\times 4'$).

We perform a spatial matching (using a matching radius of $1.5''$) between the DES and CLASH catalogues and we find 609 matched sources. Thus the DES recovered only 18\% of the sources in the CLASH catalogue. The high percentage of sources missed in DES is due to various problems, one of them being that the $griz$ 10$\sigma$ depths differ by $\gtrsim2$ mag between the two data sets. This accounts for most of the undetected sources in DES: when we simulate fake faint galaxies with \textsc{Balrog}\footnote{A software pipeline for embedding simulations into astronomical images. See: \url{https://github.com/emhuff/Balrog}.} (\citealt{suchyta}) on the DES image of RXJ2248, we find that the completeness in $riz$ bands (which are those used to run the detection) drops below 20\% between magnitude 24 and 25, justifying the incompleteness found when comparing to the even deeper CLASH survey. We also expected one of the problems to be blending, especially close to the bright cluster core. We run some completeness tests using a DES enhanced deblending catalogue (\citealt{yuanyuan}) that would increase the percentage of recovered sources to 20\%, but found that blending is not a major reason of incompleteness. Also, CLASH object detection is run on ACS+IR images, while DES detection only involves optical bands and it may miss redder sources. A visual comparison of DES and CLASH images is shown in Figure \ref{images}.

A comparison of measured isophotal magnitudes at the catalogue level between the matched galaxies in the two data sets here considered shows a mean shift $|\Delta m| \leq 0.13$ in all bands, where the offsets due to the different filters compared have been taken into account. This is true when a signal-to-noise cut $S/N>10$ is performed on the matched galaxies, and objects with saturated pixels and corrupted DES data are removed. A magnitude analysis is presented in Appendix \ref{magcomp}.

\begin{figure}
\centering
\includegraphics[width=0.48\textwidth]{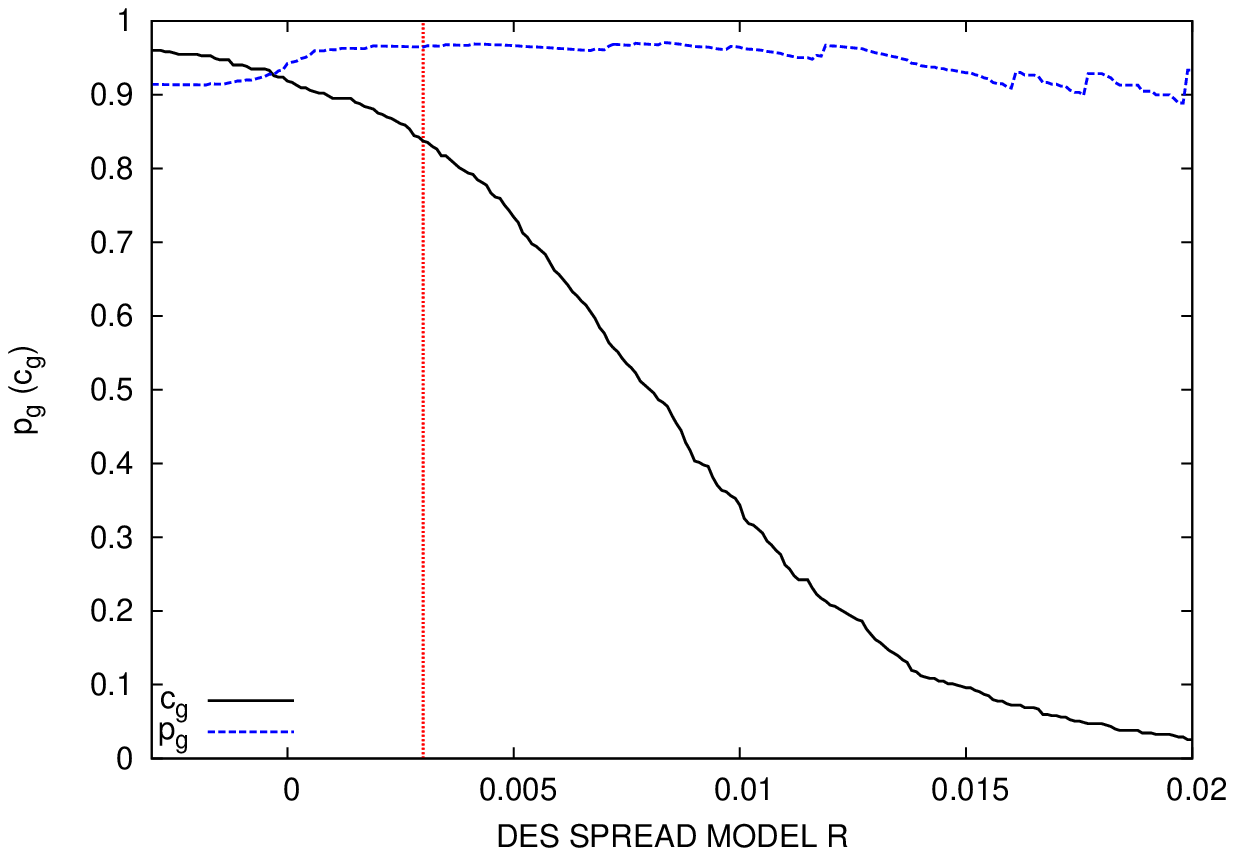}
\includegraphics[width=0.48\textwidth]{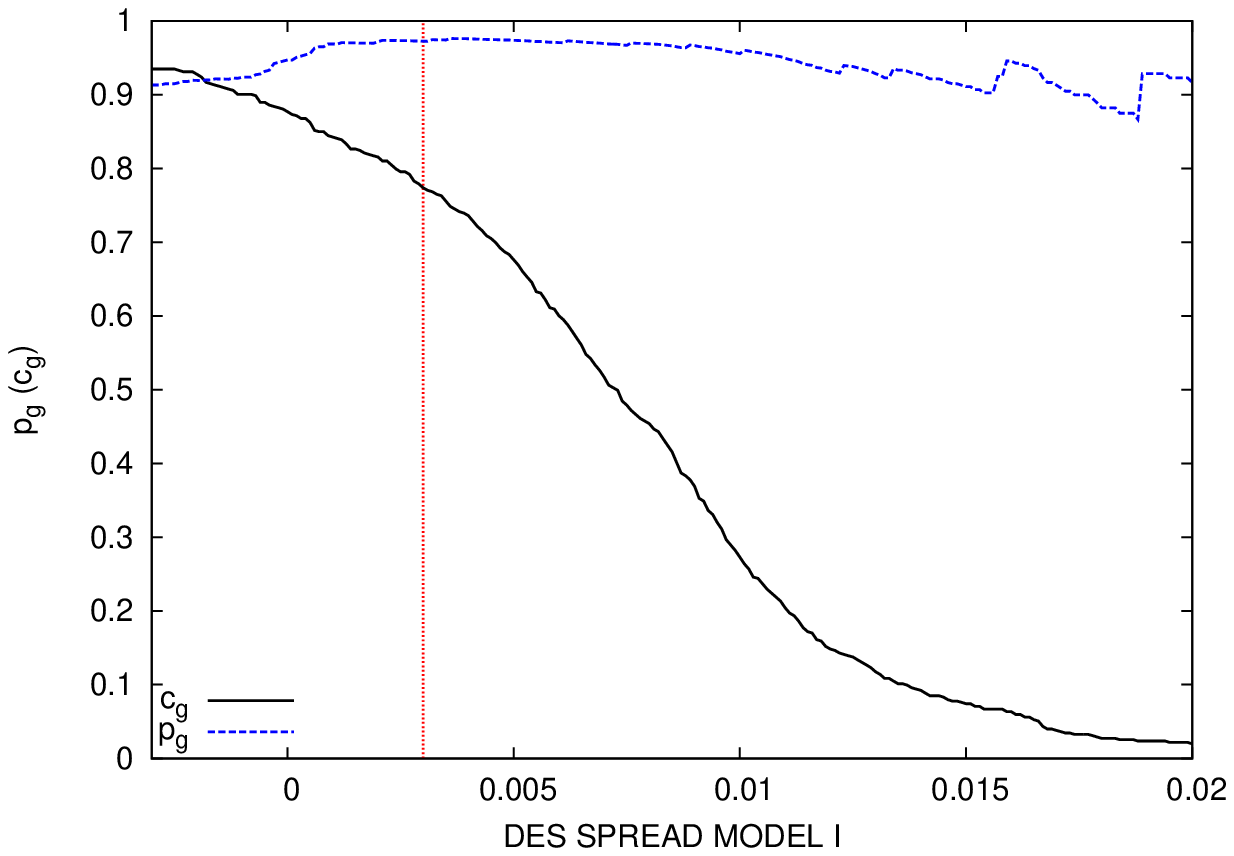}
\caption{Galaxy purity (blue dashed line) and completeness (black solid line) for the star/galaxy separation problem using the \texttt{SPREAD\_MODEL} parameter in the DES catalogue for the $r$ (top) and $i$ (bottom) bands. The red vertical line represents a typical cut used for \texttt{SPREAD\_MODEL}, which is  0.003.}\label{purity}
\end{figure}

\subsection{Star/galaxy separation}
\begin{table}\centering
\begin{tabular}{c|cccc}
$\texttt{CLASS\_STAR}$&$g$&$r$&$i$&$z$\\
\hline
 $ G({\rm DES})$&    551&      538&      535&      522\\
  $G({\rm CLASH})$ &   553 &      553&      553   &   553\\
   $S({\rm  DES})$&  58 &      71&      74&      87\\
   $S({\rm CLASH})$& 56&      56&      56&      56\\
\end{tabular}\hskip 0.3cm
\begin{tabular}{c|cccc}
 $\texttt{SPREAD\_MODEL}$&$g$&$r$&$i$&$z$\\
\hline
 $ G({\rm DES})$&    388&      463&      428&      405\\
 $G({\rm CLASH})$ &   553 &      553&      553   &   553\\
   $S({\rm DES})$&  221 &      146&     181&      204\\
  $S({\rm CLASH})$& 56&      56&      56&      56\\
\end{tabular}\caption{Number of galaxies $G$ and stars $S$ found in DES and CLASH, when considering $\texttt{CLASS\_STAR}<0.8$ (top table) and $\texttt{SPREAD\_MODEL}>0.003$ (bottom table) for galaxies in DES.}\label{numbergal}
\end{table}
For the purpose of studying the star/galaxy separation, we adopt the same notation used in \citet{soumagnac}. We study the galaxy completeness $c_g$, defined as the ratio of the number of true galaxies classified as galaxies to the total number of true galaxies (including then also the number of true galaxies classified as stars $M_G$):
\begin{equation}
c_g=\frac{N_G}{N_G+M_G}\;,
\end{equation}
where here $N_G$ is given by the galaxies in the DES catalogue, and the number of true galaxies is given by the object classified as such in CLASH.\\
Moreover the galaxy purity $p_g$ is defined as
\begin{equation}
p_g=\frac{N_G}{N_G+M_S}\;,
\end{equation}
where $M_S$ is the number of stars classified as galaxies.

We consider as true galaxies the sources that have a \textsc{SExtractor} stellarity index \texttt{CLASS\_STAR}$<0.08$ in the CLASH catalogue, otherwise they are stars. This cut has been proven to perform well in other CLASH works (e.g. \citealt{jouvel}). We try to understand if the star/galaxy performance is compatible between the two data sets.\\  

\begin{figure}\textcolor{white}{[h]}\centering
\includegraphics[width=0.5\textwidth]{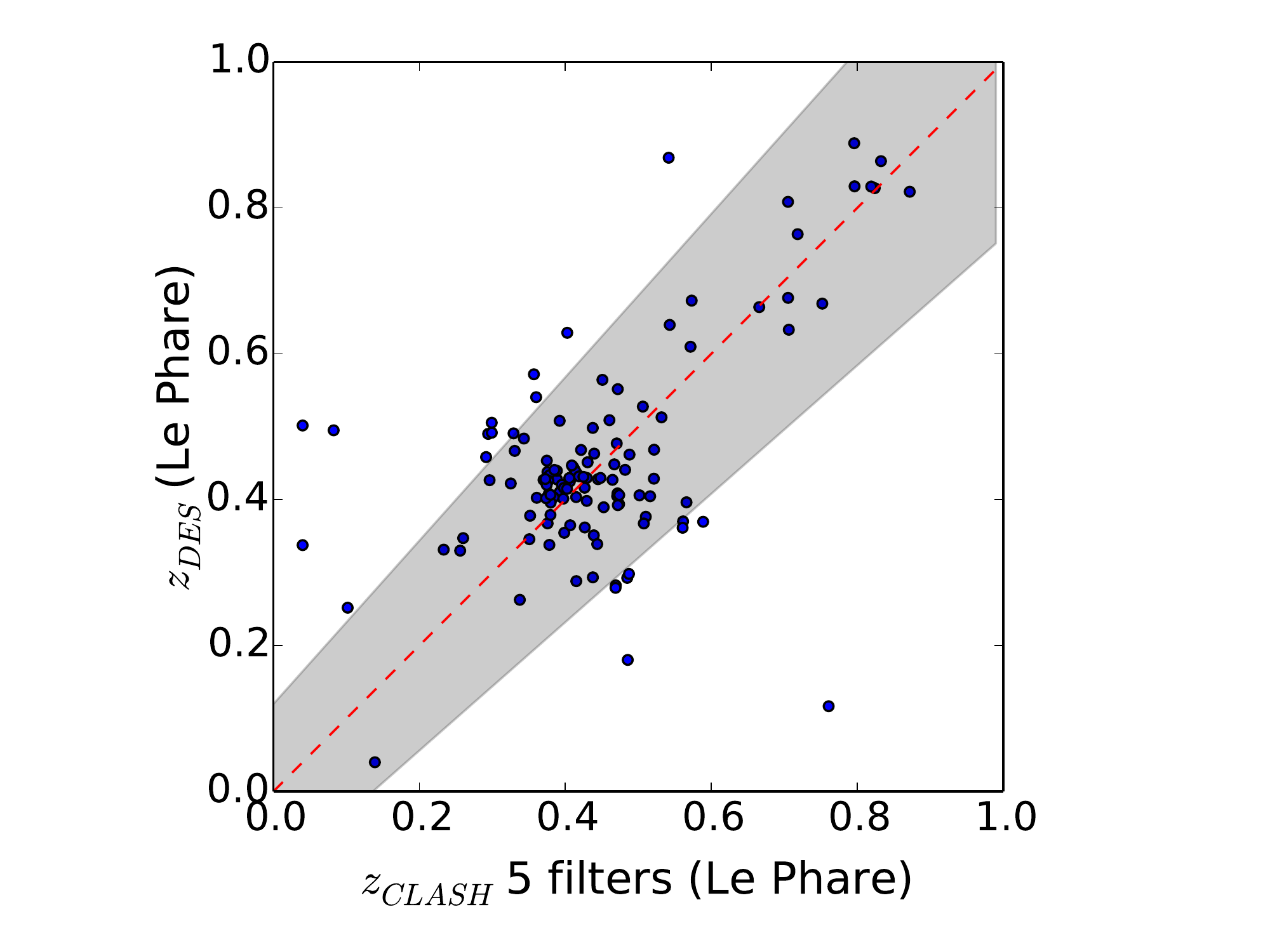}
\includegraphics[width=0.5\textwidth]{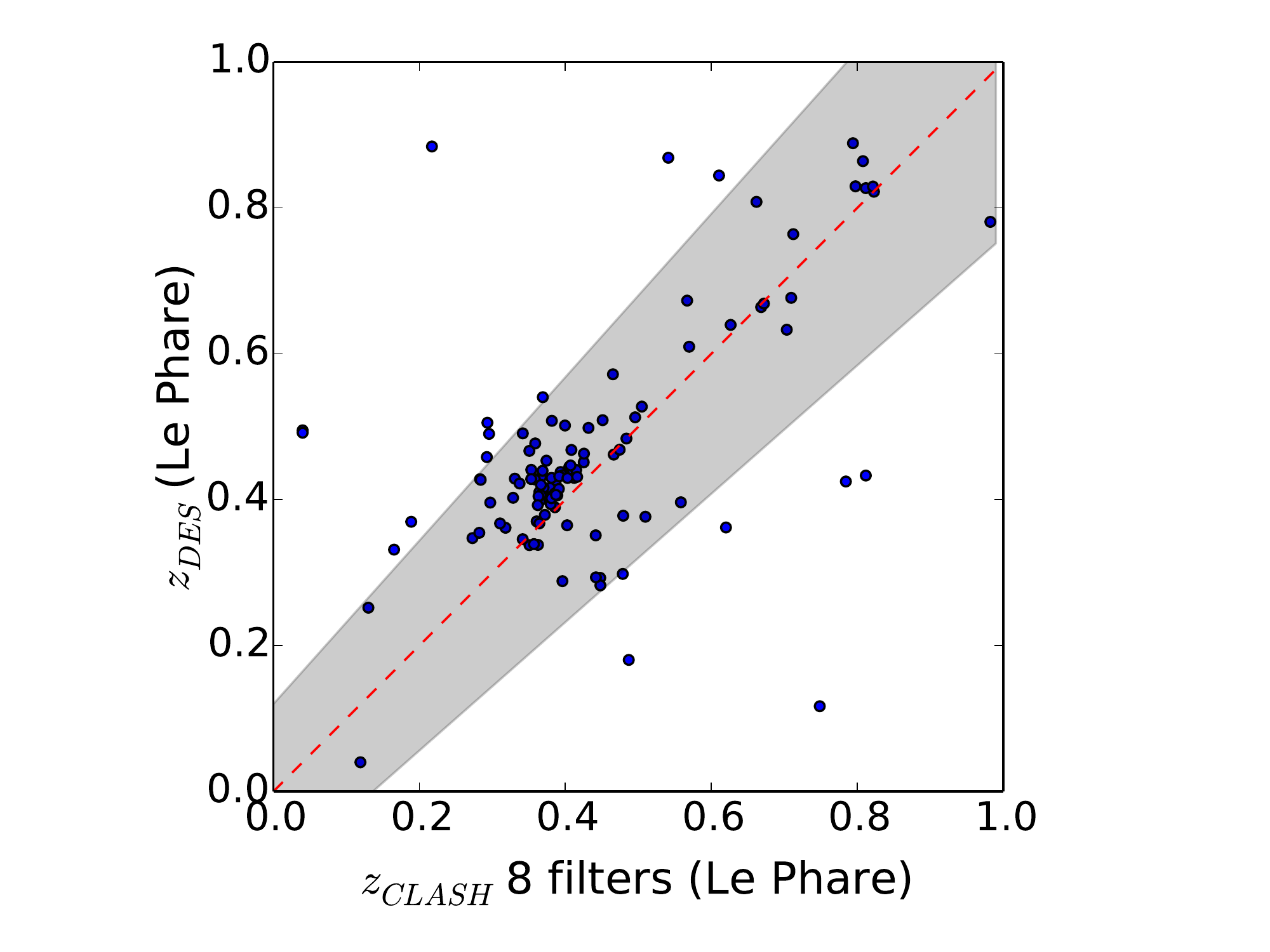}
\includegraphics[width=0.5\textwidth]{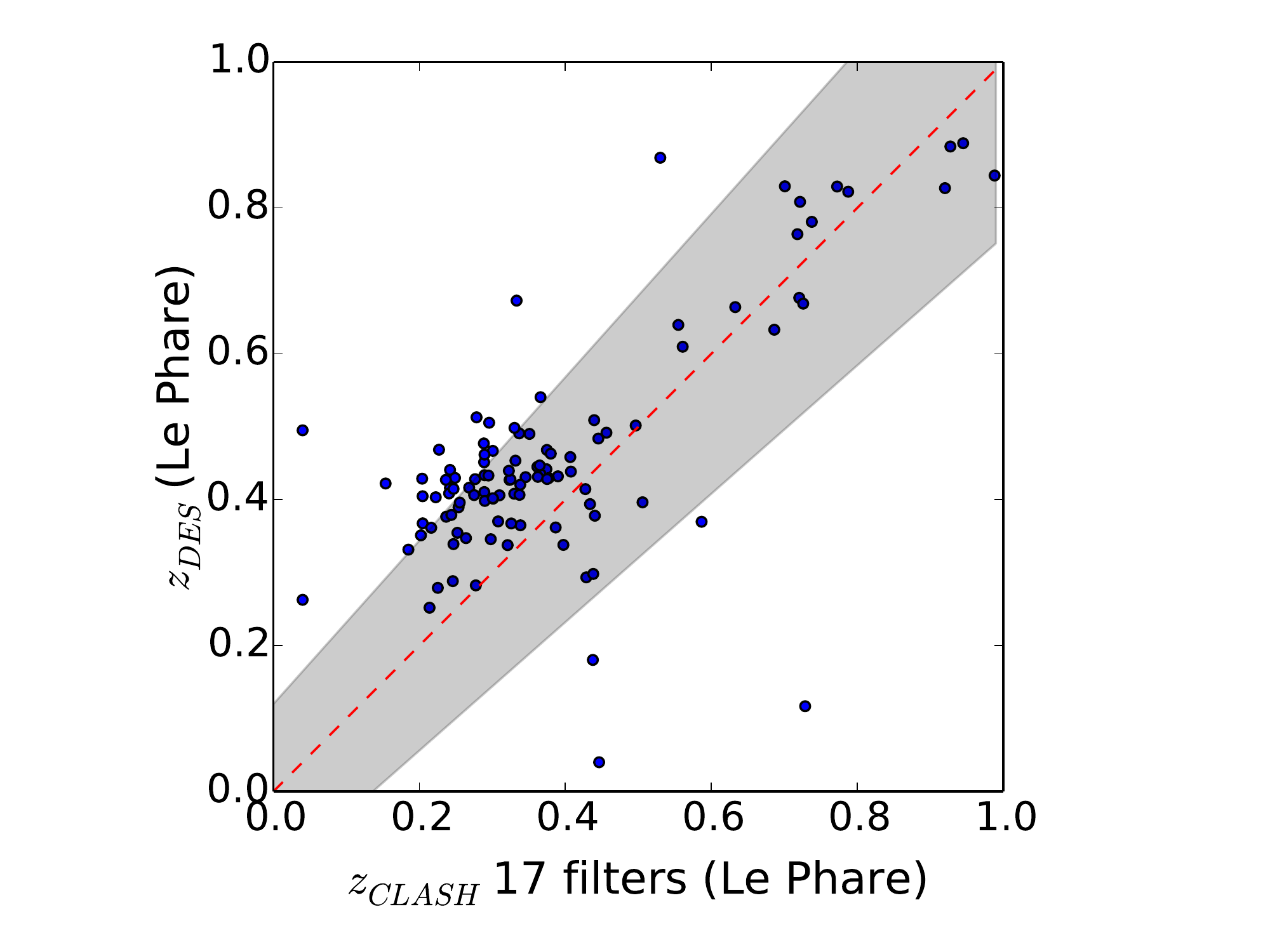}
\caption{Comparison of the DES versus CLASH photo-$z$'s for the sources matched between the two catalogues. photo-$z$'s were obtained using \textsc{LePhare}. \emph{Top:} only the 5 \emph{HST} filters similar to the $grizY$ filters in DES have been used to compute CLASH photo-$z$'s. \emph{Middle:} 3 of the \emph{HST} UV filters have been added to the 5 \emph{HST} optical filters for the CLASH photo-$z$ estimation. \emph{Bottom:} all the 17 available CLASH filters have been used to estimate the photo-$z$'s. The red dashed line represents $z_{DES}=z_{CLASH}$, and the grey area the expected DES accuracy of $|z_{DES}-z_{CLASH}|< \sigma(1+z_{CLASH})$, where $\sigma=0.12$.}\label{photoz}
\end{figure}

We first consider the \texttt{CLASS\_STAR} parameter given in the DES catalogue. We find that a cut between 0.7 and 0.9 for the \texttt{CLASS\_STAR\_I} gives purity and completeness above the 90\%. The number of galaxies and stars in the two catalogues can be found in Table \ref{numbergal}.

We also test the performance of star/galaxy separation with the \texttt{SPREAD\_MODEL} parameter (defined in \citealt{desai} and tested in \citealt{spreadmodel}). \texttt{SPREAD\_MODEL}  is a morphological star/galaxy separation parameter given by \textsc{SExtractor} which acts as a linear discriminant between the best fitting local PSF model and a slightly ``fuzzier'' version made from the same PSF model, convolved with a circular exponential model. A threshold is set to 0.003 by the DESDM pipeline to separate stars (PSF like, having absolute values below 0.003) from galaxies (non-PSF like, with values higher than 0.003). As a result, 77.4\% of the galaxies are catalogued in DES as such, and the purity is 97.3\%. 
 A plot for the purity and the completeness for varying  $\texttt{SPREAD\_MODEL\_I}$ cuts is shown in Figure \ref{purity}. We list the number of galaxies and stars in the two catalogues in Table \ref{numbergal}. It can be seen from Figure \ref{purity} that cut at lower values ($\sim 0.001-0.002$) would give a higher completeness without affecting the purity significantly. Moreover, in this case may be better using the $\texttt{SPREAD\_MODEL}$ in the $r$ band, which is deeper than the $i$ one, and this can also be seen in Figure \ref{purity}, where it is clear that, for the same cut, the completeness is higher. We also find that using the \texttt{CLASS\_STAR} parameters with the mentioned cut is more efficient than adopting the \texttt{SPREAD\_MODEL\_I} with the cut at 0.003.

\subsection{Photo-$z$}\label{sec:z}
Considering only those matched sources with a signal-to-noise ratio $S/N>10$ in the DES $i$-band, excluding stars (in this case we exclude all objects with \texttt{CLASS\_STAR\_I}$>0.8$) and objects with $\texttt{FLAGS}\neq 0$ (in order to exclude objects with saturated pixels or corrupted data, and originally blended sources) we are left with 155 sources. This is the subset of galaxies that we will use for the photo-$z$ and stellar mass comparison.

In order to estimate the photo-$z$'s, we used the publicly available software \lephare (\citealt{arnouts}, \citealt{ilbertlephare})\footnote{\url{http://www.cfht.hawaii.edu/~arnouts/LEPHARE/lephare.html}}, as it also produces the stellar masses that we want to study in this paper. Previous works on DES photo-$z$'s have tested the performance of this code in comparison with other softwares and spectroscopic redshifts. In particular, \citet{sanchez} found that \lephare fulfils the DES requirements on scatter and $2\sigma$ outlier fraction when it is run on SV data, and the metrics obtained are compatible with those from other template-based methods within 10\%. Training-based methods showed a lower bias compared to template-fitting codes, and this can be improved in the latter using adaptive recalibration methods, which are available in \textsc{LePhare}. However, stellar mass tests have not been performed with DES data so far.

We therefore need to further check the DES photo-$z$ and stellar mass estimation with \lephare first. This is where the \emph{HST} data are particularly useful in this work, as we need to check DES against a more precise photometric survey covering the wavelengths from optical to IR. 

Furthermore, \lephare is a reliable code, as seen in e.g. \citet{cosmos}.

\subsubsection{\lephare photo-$z$ technique}
 The main purpose of \lephare is to compute photometric redshifts by comparing template Spectral Energy Distributions (SEDs) to the observed broadband photometry, but it can also be used to calculate physical parameters such as stellar masses and rest-frame luminosities. Several SED sets are available within the code, and these are redshifted and integrated through the instrumental transmission curves.  Additional contribution of emission lines in the different filters can be included and extinction by dust can be taken into account.  The synthetic colours obtained from the SEDs for each redshift are then compared to the data. The best fitting template and redshift for each object is then found by $\chi^2$ minimization. In addition, prior information can be supplied, including a photo-$z$ distribution prior by galaxy type computed from the VVDS survey in the $i$ band (see \citealt{ilbertlephare}). 

\subsubsection{Results}
We run \lephare on both CLASH (with 5, 8 and all 17 filters) and DES (5 filters) catalogues, fitting the 31 synthetic SEDs templates given by the COSMOS (see \citealt{cosmos}) libraries. We use four galaxy extinction values ranging from 0.05 to 0.3 using a \citet{calzetti} extinction law for DES. For CLASH, two more extinction values are added (0.4 and 0.5), in order to take into account the wider wavelength range covered when we use all its filters. 

The results are plotted in the upper panel of Figure \ref{photoz} for the 155 matched sources when 5 filters are considered for CLASH. 
 Of all the sources considered, 85\% have a photo-$z$ which is compatible with the CLASH photo-$z$ within the DES requirement\footnote{Where $z_p$ and $z_s$ are the photometric and spectroscopic redshifts, so here we consider the CLASH photo-$z$ as the ``spectroscopic'' one. } $|z_{p}-z_{s}|< \sigma(1+z_{s})$, where $\sigma=0.12$. 
We notice an offset in the CLASH redshift when 17 filters are used (see lower panel of Figure \ref{photoz}), while 77\% of the sources still satisfy the DES requirement. This most likely stems from the inclusion of near-UV filters to get an accurate redshift from the Balmer break for galaxies below a redshift of 0.4 (see e.g. \citealt{eisenstein}). In fact, this offset starts to be seen also when adding only three UV bands (namely F336W, F390W and F435W) to the $grizY$ filters (see middle panel in Figure \ref{photoz}). A problem around redshift 0.4 for DES galaxies had already been seen in \citet{sanchez} (see their Figure 5) and \citet{bonnett}. In particular, \citet{bonnett} also pointed out a lack of matching SEDs for galaxies around redshift 0.4 with template fitting methods (see their Figure 8). 

Zero--points\footnote{Zero--points define the shift in the observed magnitudes due to various systematics.} have not been adopted in the DES photo-$z$ estimation, as we saw that their introduction causes systematic effects. Zero points are calculated using field galaxies, so we believe we would need spectroscopy in the cluster field to be helpful at photo-$z$ calibrations for this study.\\
\begin{figure}\centering
\includegraphics[width=0.5\textwidth]{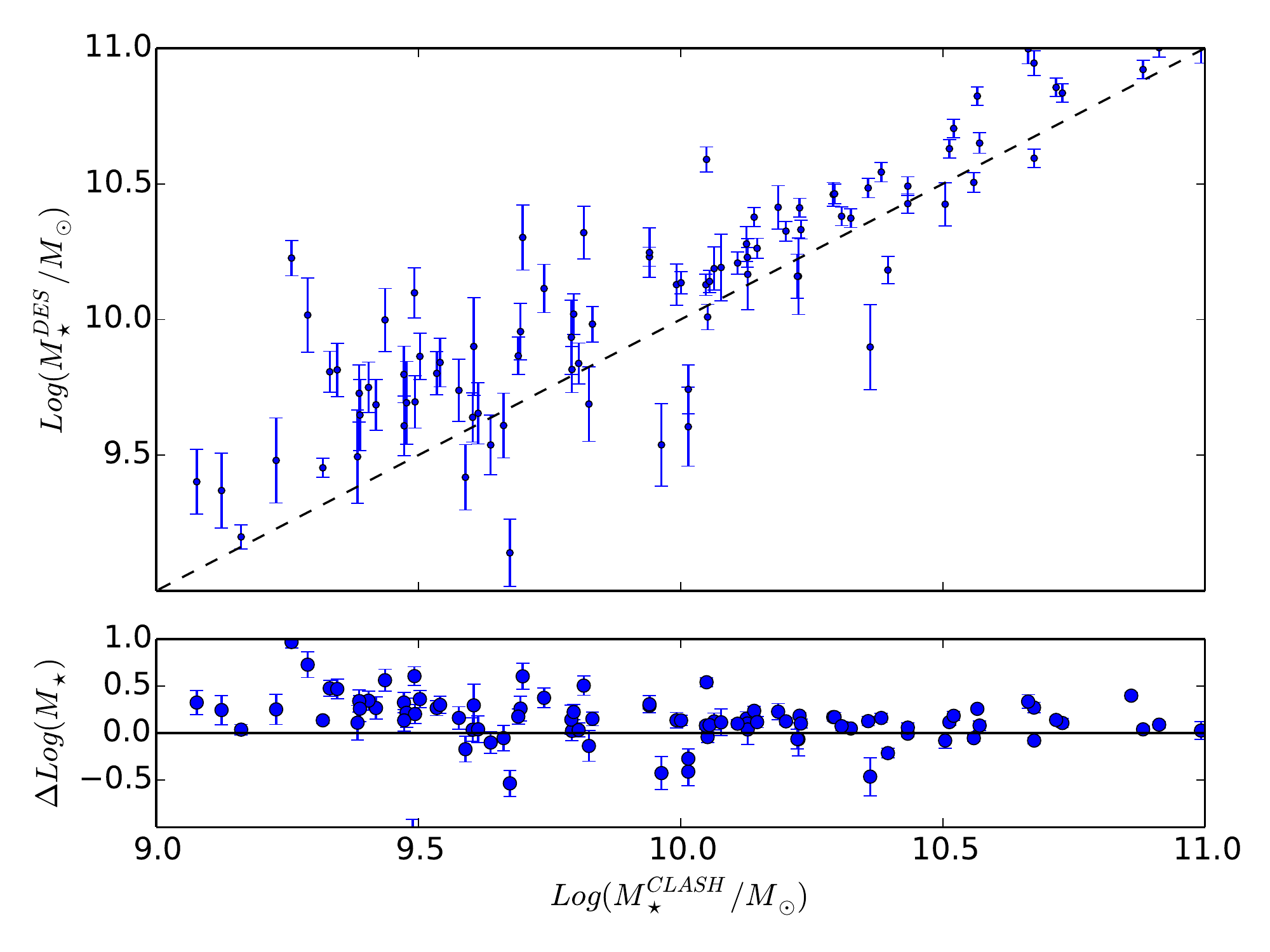}
\caption{DES stellar masses versus CLASH stellar masses computed using \textsc{LePhare}. In the stellar mass estimation, each source is  assumed to be at the redshift given as output by \textsc{LePhare}, as described in Section \ref{sec:z}. The dashed line line represents $M_\star^{DES}=M_\star^{CLASH}$. In the bottom panel $\Delta Log(M_\star)=Log(M_\star^{DES})-Log(M_\star^{CLASH})$ is presented. All available filters (\emph{i.e.} 5 for DES and 17 for CLASH) have been used in the estimation process. Uncertainties represent the 68\% Confidence Level.}\label{SMdesvsclash}
\end{figure} 
\subsection{Stellar Masses}\label{clash_sm}

Stellar masses are key observables in the study of galaxy evolutionary models. Unfortunately, they cannot be directly measured, but require multicolour photometry to be fitted with stellar population models, therefore making a series of assumptions. One of these is the galaxy redshift if spectroscopy is not available: in the view of our goal of computing the stellar mass profile of the RXJ2248 cluster, we have to bear in mind that galaxy redshift accuracy is essential not only to ensure the correct template match in the template fitting method here used and the distance to the galaxy, but also to determine the cluster membership. We will therefore see how the redshift assumptions affect the stellar mass estimation and elaborate a reasonable technique to correctly estimate the stellar mass profile.

\subsubsection{Method}
We use the same sample of matched galaxies with $S/N>10$ used in Section \ref{sec:z} and their redshift estimations in order to compute the stellar masses for both DES and CLASH using \textsc{LePhare}. In the first place, the redshifts of the galaxies are fixed to those photo-$z$'s previously computed ($i.e.$ to DES photo-$z$'s for DES stellar masses, and to CLASH photo-$z$'s for CLASH stellar masses). In the second case, we fix the galaxy redshifts at the cluster redshift for both DES and CLASH, and the \textsc{LePhare} DES photo-$z$'s are only used to select a subsample of cluster members satisfying $|z_{phot}-z_{cl}| \le 0.12$. For this subsample both DES and CLASH stellar masses are estimated.

 We chose to use \textsc{LePhare}, together with the \citet{bc} templates, as this combination has been shown to be robust in the estimation of physical parameters of galaxies (\citealt{ilbert2010}).

\begin{figure}\centering
\includegraphics[width=0.5\textwidth]{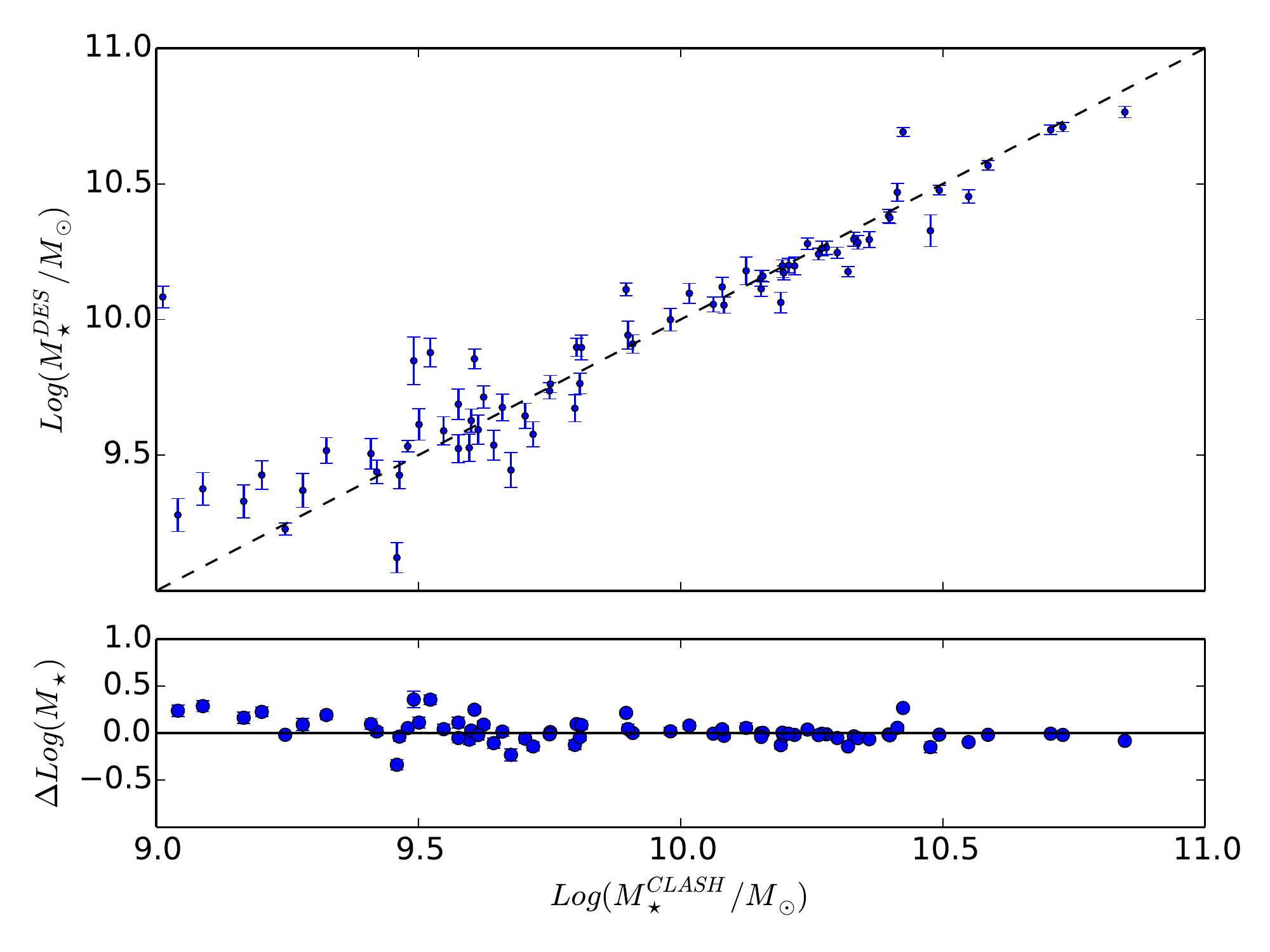}
\caption{DES stellar masses versus CLASH stellar masses computed using \textsc{LePhare}. Here only sources around the cluster redshift are considered (\emph{i.e.} sources with a DES photo-$z$ that satisfies $|z-z_{cl}| \le 0.12$, where $z_{cl}=0.3475$ is the cluster redshift). In the DES and CLASH stellar mass estimation, these galaxies are all assumed to be at $z_{cl}$. The dashed line line represents $M_\star^{DES}=M_\star^{CLASH}$. In the bottom panel $\Delta Log(M_\star)=Log(M_\star^{DES})-Log(M_\star^{CLASH})$ is presented. The offset seen in Figure \ref{SMdesvsclash} seems to disappear in this plot, showing that this effect was due to the photo-$z$ offset. All available filters (\emph{i.e.} 5 for DES and 17 for CLASH) have been used in the estimation process. Uncertainties represent the 68\% Confidence Level.}\label{SMdesvsclash_zcl}
\end{figure}
We derive our stellar mass estimates by fitting synthetic SEDs templates while keeping the redshift fixed as described previously in the two cases. The SED templates are based on the stellar population synthesis (SPS) package developed by \citealt{bc} (BC03) assuming a \citet{ch} initial mass function (IMF). Our initial set of templates includes 9 models using one metallicity ($Z=1 Z_\odot$) 
and nine exponentially decreasing star formation rates $\propto {\rm e}^{-t/\tau}$ where $t$ is the time and $\tau$ takes the values $\tau = 0.1,0.3,1,2,3,5,10,15,30$ Gyr. The final template set is then generated over 57 starburst ages ranging from 0.01 to 13.5 Gyr, and four extinction values ranging from 0.05 to 0.3 using a \citet{calzetti} extinction law. For CLASH, two more extinction values are added (0.4 and 0.5).

The uncertainties on our stellar masses estimates (\texttt{MASS\_BEST} from \textsc{LePhare}) are given by the 68\% confidence limits on the SED fit.

\begin{figure}\hskip -0.7cm
\includegraphics[width=0.6\textwidth]{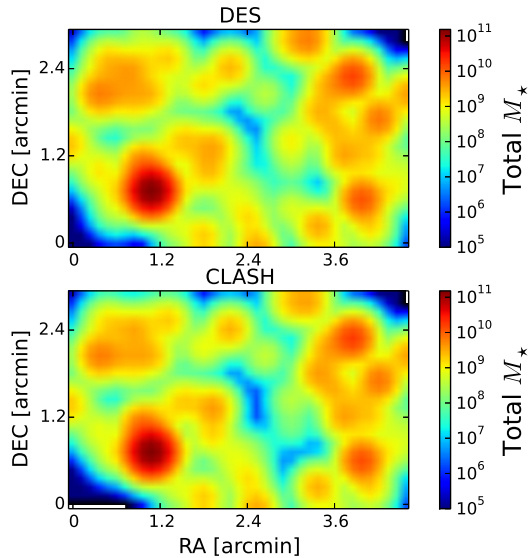}\caption{DES and CLASH total stellar mass maps computed using \lephare for the galaxies matched between the two catalogues. The stellar masses plotted are the same as those shown in the bottom panel of Figure \protect\ref{SMdesvsclash_zcl} (\emph{i.e.} all the galaxies with a DES photo-$z$ that satisfies $|z-z_{cl}| \le 0.12$, where $z_{cl}$ is the cluster redshift, have a redshift fixed to $z_{cl}$ in the SED fitting). The map is centred on the BCG, but its stellar mass is not visible as it was originally blended and therefore did not pass the quality flag cut applied in Section \ref{clash_sm}. The resolution is $0.12'/$pixel and the map is smoothed with a Gaussian of $\sigma=0.144$ arcmin. At the cluster redshift, $1'$ corresponds to 294 kpc in the assumed cosmology.}\label{desclashmaps}
\end{figure}

\subsubsection{Results}
 In Figure \ref{SMdesvsclash}, we show the comparison between DES and CLASH stellar mass estimates for the first case, where the redshifts are fixed to the \textsc{LePhare} estimates.
 The linear correlation between the two estimates is clear, but there is an offset of mean value $\sim 0.16$ dex. This should be considered in light of two aspects:
\begin{enumerate}
\item the offset in the photo-$z$'s that we addressed in Section \ref{sec:z};
\item the uncertainties in the DES stellar masses may be underestimated as those are the 68\% confidence limits on the SED fit and do not take into account systematic error contribution.
\end{enumerate}

In Figure \ref{SMdesvsclash_zcl} we show the results for the second case, where we select the galaxies with a DES photo-$z$ close to the spectroscopic cluster redshift $z_{cl}$, satisfying $z_{cl}-0.12<z_{phot}<z_{cl}+0.12$. The redshift of these sources is fixed at $z_{cl}$ in the stellar mass estimation, and the reason for this choice is twofold:
\begin{enumerate}
\item to minimize circularity associated with using \textsc{Le Phare} to both measure redshifts and stellar masses;
\item to take into account the shift in the redshift estimates pointed out in Section \ref{sec:z} (and therefore put at the correct cluster redshift the cluster members which photo-$z$ appeared to be at $z_{phot}\sim 0.4$).
\end{enumerate}

Of course this choice results in considering some sources as being at $z_{cl}$ even though they are not, and we shall take this into account in the following.  A consistency test for the choice of redshifts done is presented in Appendix \ref{test}.

The correlation between the estimated stellar masses significantly increases if also the CLASH sources are set to be at the cluster redshift, as seen in a comparison of Figure \ref{SMdesvsclash} with Figure \ref{SMdesvsclash_zcl}, where we find that stellar masses from DES can be estimated within 25\% of CLASH values. This shows that the offset seen in the former is due to the offset in the redshifts given in input, rather than other systematics. Therefore, the wavelengths covered by the DES broadband filters are capable of providing a good estimation of stellar mass if the photometric redshift is sufficiently precise. 

In Figure \ref{desclashmaps} we show the spatial distribution of the total stellar mass spatial distributions for both DES and CLASH in the CLASH field ($\sim 4.8'\times 4.2'$), represented with a resolution of $0.12'/$pixel and smoothed with a Gaussian of $\sigma=0.144$ arcmin. The stellar masses of galaxies are summed over in each pixel.
Obviously, the two samples show very good agreement in terms of the spatial distribution of stellar mass. The Pearson coefficient for the pixel-by-pixel stellar mass values of the two non-smoothed maps is 0.93. The difference map without any smoothing has a mean of 0.02 and $\sigma=0.12$.

\section{Dark Matter and Stellar Masses}\label{DMandSM}
In this section we study the stellar mass radial profile of the cluster and relate it to that of dark matter that has been obtained though DES weak lensing studies in \citet{melchior}. As shown in Section \ref{clash_sm}, stellar mass estimation can be biased if the redshift assumed is biased too: this is a problem if one wants to compute stellar masses with photometric surveys data. In the lack of spectroscopy, it is a common method to perform estimation of the photometric redshift and SED fitting in two steps, which involve fixing the redshift of a galaxy at the best fitting value obtained in the first step. Although this may not be the most elegant way of solving the problem, it has been proven to lead to only a small bias in the SED fitting parameters, when compared to results given by the simultaneous estimation of redshift and stellar mass (see \citealt{acquaviva}). Nevertheless, here we want to adopt a more consistent methodology for the stellar mass estimation, taking advantage of the fact that we are looking at a cluster with a known redshift. This technique is outlined in the first part of this section, followed by a study of the different mass radial profiles obtained. To allow a straightforward comparison with the weak lensing reconstructed mass, we compute total stellar mass and surface density on a projected 2-dimensional plane, \emph{i.e.}:
\begin{equation}
M_\star(R)=\sum_i m_\star^i \qquad\Sigma_\star(R)=\frac{\sum_i m_\star^i}{A_\mathrm{annulus}}\, ,
\end{equation}
where the sums are intended over the galaxies within annuli of projected radius $R$. Similar definitions apply for the cumulative distributions $M_\star(<R)$ and $\Sigma_\star(<R)$, computed within circles of radius $R$.  The centre of the image is taken to be that of the BCG. At last, we present a comparison between the stellar and total DES mass maps.

\subsection{Galaxy samples and stellar mass estimates}   
Our goal is to compare the reconstructed mass from weak lensing to the total stellar mass of the cluster members. In order to do so, we split the galaxies into two populations. The following steps are performed:
\begin{itemize}
\item We select the red members using the \textsc{RedMaPPer} SV catalogue (\citealt{rykoff}), which identifies cluster members with high precision (\citealt{rozo}). Their stellar masses are computed using the same parameters presented in Section \ref{clash_sm} (but fixing the redshift at $z_{cl}$). 
\item The \textsc{RedMaPPer} galaxies profile has been corrected by a factor representing the contribution coming from faint sources at luminosities smaller than the limit of the sample (0.2$L^*$). This is done by integrating the luminosity with a Schechter function, \emph{i.e.} we computed the fraction:
\begin{equation}data set
F_L=\frac{\int^{0.2L^*}_0 L \phi(L)d L }{\int_{0}^\infty L\phi(L)d L} \, ,
\end{equation}
where $\phi(L)=\phi_*(\frac{L}{L^*})^\alpha {\rm e}^{-L/L^*}$ with $\alpha =-1$ [as done in \citet{redpaper} for the SDSS sample, that has properties similar to DES]. We find that the galaxies below the luminosity limit contribute to a fraction $F_L=0.18$ of the total luminosity, and therefore, assuming a constant $M_\star/L$ ratio for the red galaxy population, they contribute to the same percentage of stellar mass.
\item The contribution to the total stellar mass of each red member is weighted by its membership probability (reported in the RedMaPPer catalogue).
\item In order to study the mass profile at radii higher than $r_{200c}$, we decide not to neglect the contribution coming from the bluer population. First, we exclude all objects with saturated pixels or corrupted data, but include galaxies that were initially blended (such as the BCG). Then we select the rest of the galaxies in the field of view that have magnitudes $m_i$ in the $i$ band satisfying $m_i^\mathrm{BCG}<m_i<m_i^\mathrm{lim}$. In this way we exclude any source which is brighter than the BCG and cut at $m_i^\mathrm{lim}=21$ mag in order to ensure the completeness of the sample. After having performed a SED fitting as in the previous step, we filter out all galaxies that do not give a good fit (cutting on reduced $\chi^2<2$) when the redshift is fixed at the cluster value. 
\end{itemize}

\subsection{Masking and background correction}

We estimate the survey area lost due to masked regions and blending of faint galaxies with large cluster members near the core. We calculate corrections for both effects as follows.
\begin{itemize}
\item \textsc{Healpix} \citep{healpix} maps of depth and masking fractions are produced for DES with \textsc{Mangle}
\citep{mangle}. From these, we calculate mean depth and fractions of masked area in our set of annuli. The depth is approximately constant out to $\approx 50$ arcmin from the BCG, which defines the outer limit of the area used for our background estimation scheme. Masking fractions are below 5 per cent for all annuli and applied to the binned stellar mass estimates from both galaxy samples.
\item For the blue galaxy sample, some objects are lost due to blending with cluster member galaxies. Without correction, this would bias our stellar mass estimates of blue galaxies near the cluster centre low. We estimate the area lost in each annulus as the isophotal area above the \textsc{SExtractor} detection threshold,
\texttt{ISOAREA\_I}. This yields a $\approx 7$ per cent correction in the innermost arcminute, which drops quickly towards larger radii. For the blue galaxy sample, this correction and the masking fraction are applied in an additive fashion.
\end{itemize}

The contribution coming from galaxies that do not actually sit at the cluster redshift is removed from the blue galaxies sample by performing a background subtraction: we estimate the projected surface density of the stellar mass $\Sigma_\star(R)$ at large radii ($30-50$ arcmin, which means outside $\approx 4r_{200c}$ \footnote{$r_{200c}=2200$ kpc from the NFW fit of \citet{melchior}, which means $r_{200c}\simeq7.46'$ at the cluster redshift}), where the stellar mass profile tends to become flat. The value found is $\Sigma_\star=1.36\times 10^{10} M_\odot/\textrm{arcmin}^2$ and this is subtracted on the smaller scales, with an uncertainty given by a Poissonian error. The remaining stellar mass contribution is then added to that of the red galaxies.

\subsection{Stellar mass profile}

\begin{figure*}
\includegraphics[width=1\textwidth]{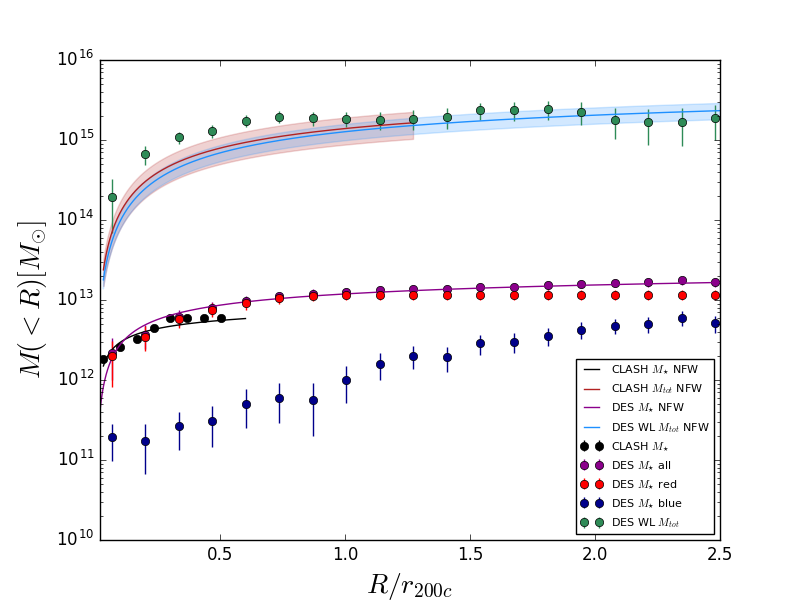}
\caption{Cumulative radial distributions of total stellar mass derived in this paper for the DES red (red points), blue (blue points) and all galaxies (purple points) in the cluster, together with the total, non-parametric mass profile reconstructed from DES weak lensing (\citealt{melchior}, green points). The purple solid line is our NFW fit to the DES stellar mass profile, while the blue solid line is the NFW best fit from DES weak lensing. The black points represent the CLASH total stellar mass profile computed in this paper, with our NFW fit (black solid line). The red solid line is the \citet{umetsu15} NFW best-fit for CLASH from a strong lensing, weak lensing and magnification joint analysis. This profile is restricted to the NFW fitting range $R<2 {\rm Mpc }\; h^{-1}$ chosen in \citet{umetsu15}, which is larger than the \emph{HST} field of view as other data sets were used in a joint analysis. The radius $R$ is projected, and $r_{200c}=2.2$ Mpc. Errorbars show the 68\% confidence level. }\label{sm_profile}
\end{figure*}

We look at the radial distribution of stellar mass, taking into account both the red cluster members present in the \textsc{RedMaPPer} catalogue, and the blue members, as explained in the previous section. The splitting into red and blue galaxies is justified by the possibility of improving the SED fitting by using different priors for the two populations, and considering the systematics differently. In fact, it is well known that stellar masses estimated for quiescent galaxies are more reliable than for star-forming ones, partially because the colour--$M_\star/L$ (from which $M_\star$ is derived) relation is more uncertain for very blue colours (see e.g. \citealt{conroy} ; \citealt{banerji}).

The total stellar mass cumulative profiles $M_\star(<R)$ for the red and blue galaxies are shown in Figure \ref{sm_profile}. Within the innermost 5 arcmin, the contribution of the red cluster members to the total stellar mass is dominant ($\gtrsim 80\%$) with respect to the bluer galaxies, while at larger radii, namely outside $r_{200c}$, the second population considered gives a $20\ldots 50\%$ contribution to the total stellar mass.
In Figure \ref{sm_profile} we also plot the stellar mass profile from CLASH, where the galaxy cluster members were selected cutting on the CLASH photometric redshift with $|z_{phot}-z_{cl}|<0.12$.
 
 \subsection{Comparison to total mass from weak lensing}
 
The weak lensing mass profile is computed through the aperture mass densitometry (see \citealt{clowe}) using $M_{tot}(<R)=\upi R^2\zeta(<R)\Sigma_{cr}(z_l,z_s)$, where $\zeta(<R)=\bar{\kappa}(<R)-\bar{\kappa}(r_1<r<r_{2})$ is the difference between the mean convergence within a circular aperture of radius $R$ and the mean convergence between $r_1$ and $r_2$ (annulus radii that are fixed for all the apertures in the measurement), $z_l$ and $z_s$ are the redshift of lens and sources.  The convergence $\kappa$ is defined as the projected surface mass density  $\Sigma$, in units of the critical surface mass density $\Sigma_c$:
\begin{equation}\centering
\kappa=\frac{\Sigma}{\Sigma_c}\, , \qquad \Sigma_c=\frac{c^2}{4\upi G}\frac{D_s}{D_d D_{ds}}\,
\end{equation}
where $D$ stands for angular diameter distance and the subscripts $s, d, ds$ indicate the distance from the observer to the source, from the observer to the lens, and from the lens to the source respectively. In particular, \citet{melchior} used $r_1=30$ arcmin and $r_{2}=45$ arcmin. This explains our choice of estimating the stellar mass surface density background in the range 30 to 50 arcmin, where its profile is also essentially flat.
In Figure \ref{sm_profile} we also present the NFW mass profile derived by using the best fit parameters as found in \citet{melchior} for this cluster.
Given the similarity between the WL and  stellar mass profiles, we try to fit the stellar mass one with a NFW projected mass profile, as  the one derived in e.g. \citet{nfwfit}:

\begin{equation}
M(<x) = \left\{ \begin{array}{ll}
\frac{3 \delta_c M_{200c}}{200 c_{200}^3}
\left[\frac{2}{\sqrt{1-x^2}}{\rm arctanh}\sqrt{\frac{1-x}{1+x}}+ {\rm ln}(\frac{x}{2})
 \right] &\\
 \hspace{4.2cm} \mbox{$\left({\rm if }\;x < 1\right)$} &\\ 
 & \\
\frac{3 \delta_c M_{200c}}{200 c_{200}^3}\left[ 1+{\rm ln}(\frac{1}{2})
 \right]  & \\
   \hspace{4.2cm}\mbox{$\left({\rm if }\;x = 1\right)$}&\\ 
 & \\
\frac{3 \delta_c M_{200c}}{200 c_{200}^3}\left[\frac{2}{\sqrt{x^2-1}}{\rm arctan}\sqrt{\frac{x-1}{1+x}}+ {\rm ln}(\frac{x}{2}) \right]
&\\
 \hspace{4.2cm} \mbox{$\left({\rm if }\;x > 1\right)$} &
\end{array}
\right.\label{2dmass}
\end{equation}
where $x=R/r_{s}$, $c_{200}=r_{200c}/r_s$ is the concentration parameter and 
\begin{equation}
\delta_c=\frac{200}{3}\frac{c_{200}^3}{{\rm ln}(1+c_{200})-c_{200}/(1+c_{200})}\,.
\end{equation}

Our non-linear least squares fit uses the Levenberg-Marquardt algorithm and gives the following parameters for DES stellar mass profile: $M_{200c}^\star= (5.38\pm 0.11)\times 10^{12} M_\odot$, $c_{200}^\star= 2.4  \pm 0.13 $ with a reduced $\chi ^2=0.6$.

\begin{figure}
\includegraphics[width=0.5\textwidth]{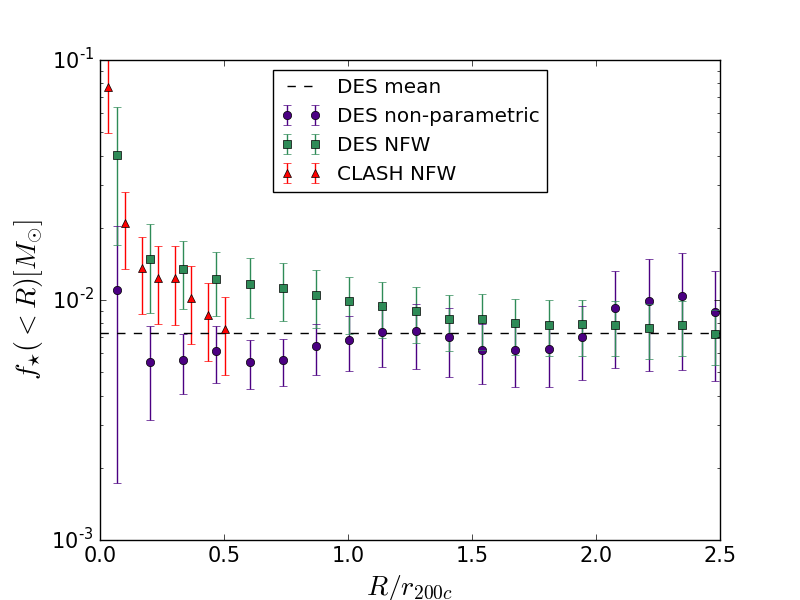}
\caption{Cumulative radial distribution of the fraction of stellar mass of the galaxies in the cluster as computed in this work over the total mass from lensing studies. For the DES $f_\star$, the total mass comes from the non-parametric reconstruction of the weak lensing shear profile done in \citealt{melchior} (purple points), or from their NFW best fit (green squares). For CLASH, $M_{tot}$ is a result of the \citet{umetsu15} NFW best-fit for CLASH from a strong lensing, weak lensing and magnification joint analysis (red triangles). The mean DES stellar mass fraction from non-parametric weak lensing mass profile is $f_\star=(7.3 \pm  1.7)\times10^{-3}$, and is represented by the dashed line. The radius $R$ is projected, and rescaled with $r_{200c}=2.2$ Mpc. Errorbars show the 68\% confidence level.}\label{f_profile}
\end{figure}

From the DES stellar mass profile and the aperture mass densitometry total matter profile, we derive the stellar mass fraction $f_\star (<R)=M_\star(<R)/M_{tot}(<R)$, which is represented by the purple points in Figure \ref{f_profile}. Within $r_{200c}$ radius, we find $f^{DES}_\star (<r_{200c})=(6.8\pm 1.7)\times10^{-3}$, compatible within 1$\sigma$ in the outer regions with the result from \citet{bahcall}: $f_\star \simeq (1.0 \pm 0.4)\times10^{-2}$ above $\sim 300 h^{-1}$ kpc. In their paper, \citet{bahcall} examine the stellar fraction profile by stacking $> 10^5$ SDSS groups and clusters, divided into 3 richness subsamples.\footnote{They define the richness $N_{200}$ as the number of galaxies in the red sequence with rest-frame $i$-band luminosity $L_i>0.4L^*$ located within a radius $r^{gals}_{200}$ from the BCG (i.e. within the radius where the local galaxy overdensity is 200).} Inside $r_{200c}$ we recover a lower stellar mass fraction compared to their work. The discrepancy can be explained in light of the different analyses carried out in \citet{bahcall}:
\begin{itemize}
\item \citet{bahcall} stack clusters with different properties and at different redshifts.
\item They included the contribution of the diffuse intracluster light (ICL), which increases $f_\star$ by a factor of 1.15 within $r_{200c}$.
\item The luminosity profiles and weak lensing mass profiles have been de-projected to obtain 3D profiles in their work. On the other hand, considering the projected $f_\star$ means that we are including the contribution of the cluster outskirts along the line of sight when we look at cluster core. In these regions, the stellar mass fraction is lower, and this tends to reduce 2D $f_\star$ at small radii with respect the 3D behaviour.
\end{itemize}

 On the other hand, the average stellar mass of the Universe, estimated to be $f_{\star,cosmic}=(9\pm1)\times 10^{-3}$ (as derived in \citealt{bahcall}) is recovered outside $r_{200c}$, as we would expect even for a projected profile.
 
Overall, no particular radial trend is found, in agreement with \citet{bahcall} and also with \citet{andreon}, who studied the stellar-to-total mass ratio of three CLASH clusters at $z\sim 0.45$. Nevertheless, a radially varying profile might be hidden by the large errors. In order to reduce the latter, dominated by the weak lensing reconstructed mass, and have a precise estimation of the stellar mass fraction, we will need to apply the same reasoning to a large sample of DES clusters.

If we take the NFW mass profile with the lensing best-fit parameters as total mass in $f_{\star}$, we get the green points in Figure \ref{f_profile} for DES, and the red ones for CLASH. Towards the centre of the cluster these profiles are higher than the one previously discussed. This is due to the fact that in this case $M_{tot}$, as  can be seen in equation (\ref{2dmass}), goes to zero for $R \to 0$, while the BCG stellar mass contributes to $M_\star$ up to very small radii. Moreover, the halo/cusp problem (see e.g. \citealt{cusp}) is a known problem of the NFW profile, which will therefore produce different results from a non-parametric mass profile from weak lensing. Use of the same dark matter halo parameterization brings the two data sets into agreement at the $1\sigma$ level.

\subsection{DES stellar masses and weak lensing mass maps}\label{sec:maps}
In this section, we explore the correlation between the stellar mass maps and the DES weak lensing mass map by \citet{melchior}.
\citet{melchior} adopted the aperture-mass technique from \citet{schneider1996} and computed the map of $M_{ap}/\sigma_{ap}$ for this cluster, where $M_{ap}$ is the aperture mass and it is defined as the sum of all ellipticity measurements $\epsilon_t(\vartheta_j)$ inside a circular aperture:
\begin{equation}
M_{ap}(\vartheta)=\sum_jQ(|\vartheta - \vartheta_j|)\epsilon_t(\vartheta_j)\;.\label{mapeq}
\end{equation}

In \reff{mapeq} $Q$ is a weight function, and the ellipticities are computed with respect to the centre $\vartheta$ of the circular aperture. The variance of the aperture mass is given by:
\begin{equation}
\sigma^2_{M_{ap}}=\frac{\sigma_{\epsilon}^2}{2}\sum_jQ^2(|\vartheta - \vartheta_j|)\;.
\end{equation}

\begin{figure}
\includegraphics[width=0.55\textwidth]{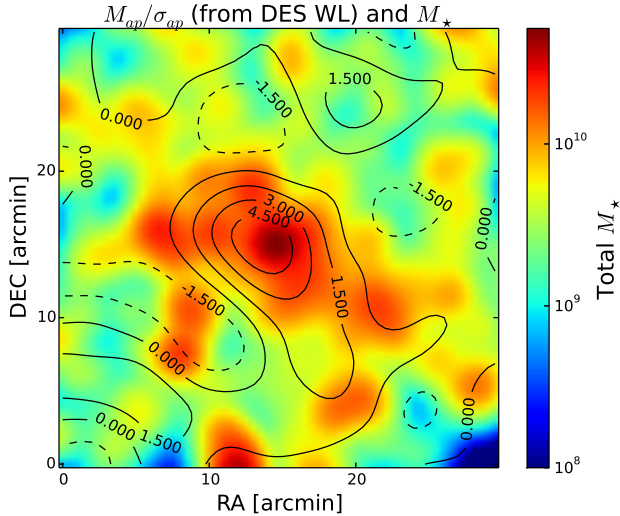}
\caption{DES total stellar mass distribution (coloured density plot) compared to the mass map (\emph{i.e.} map of $M_{ap}/\sigma_{ap}$, in contours) from the weak lensing analysis by \protect\citet{melchior}. Both maps have a resolution of $0.4997'/$pixel and have been smoothed with a Gaussian of $\sigma = 1'$.}\label{densitymapsWL}
\end{figure}

In Figure \ref{densitymapsWL} we show the DES aperture mass map (black contours) and our stellar mass map (coloured density map) in $30'\times 30'$ around the BCG position. Both maps, have a resolution of $0.4997'/$pixel and have been smoothed with a Gaussian of $\sigma = 1'$.\\

An elongated structure spanning for $\sim 4$ Mpc around the BCG is clearly present in both maps, as well as a few clumps lying inside and outside the $r_{200c}$ radius. Note that the mass structures that can be seen in the total mass map may lie outside the cluster but still cause the lensing, as they are along the line of sight, while the DES galaxies considered here, are only those at the cluster redshift. This fact partially explains why the peaks and minima in the stellar mass and aperture mass maps may be not always coincident.  Also, the peak of the stellar mass distribution coincides with the BCG position, while the weak lensing map shows a small offset of the peak from the BCG: \citet{gruen} already addressed this effect to the expected shape noise studied by \citet{dietrich}.

The correlation that can be seen by eye between $M_{ap}/\sigma_{ap}$ and $\log{M_\star}$, is quantified by a Pearson coefficient of $r=0.30$  when cross-correlating the maps pixel by pixel. Again, the correlation expected between stellar mass and total matter (quantified in this case by the aperture mass and convergence) in a cluster is diminished by the fact that the gravitational lensing gives an integrated information about the mass along the line of sight. 

\newpage
\section{Conclusions}
We compared the catalogues derived from DES and CLASH observations of the galaxy cluster RXC J2248.7--4431, treating CLASH as a validation set for DES. Photometric redshifts and stellar masses for both data sets were computed using \textsc{LePhare}, and we found that stellar masses can be estimated with good precision with DES, despite the lower number of bands available. Gravitational lensing results from both DES and CLASH were used to compare stellar and total mass maps, as well as the mass profiles and stellar mass fraction. 

Following are the conclusions.
\begin{enumerate}
\item \emph{HST} data can be used as a validation set for DES data and results. We expect 3 more CLASH clusters (Abell 383, MACS0416.1--2403 and MACS0429.6--0253) to be included in the whole DES footprint.
We found that in this case, using the \texttt{CLASS\_STAR} parameters with the mentioned cut is more efficient than adopting the \texttt{SPREAD\_MODEL\_I} with the cut at 0.003.
\item DES photo-$z$'s are compatible with the 17 \emph{HST} filters photo-$z$'s within the DES requirements. The $z\sim0.1$ offset observed in the DES photo-$z$'s, is due to a colour-redshift degeneracy that cannot be broken without UV bands at redshifts below 0.4. We found that such offset would percolate into the stellar mass estimates and bias the results by $\sim 0.16$ dex. In order to perform stellar mass studies, we therefore overcame the problem of the redshift estimation for single cluster members by devising a technique as follows. This method separately treats the red galaxies, as found by \textsc{RedMaPPer}, and the blue galaxies. The redshift information can either be the spectroscopic redshift of the cluster, if available, or the \textsc{RedMaPPer} cluster photometric redshift (as \textsc{RedMaPPer} photo-$z$'s are estimated with high precision). In order to estimate the blue galaxies contribution to the stellar mass, we perform a background subtraction which is only possible thanks to the DECam wide field of view. 
\item We then estimated total stellar mass and stellar mass fraction profiles for both DES and CLASH, reaching large radii with DES. Within the projected $r_{200c}$ radius, we find a fraction of stellar mass over total mass (derived from weak lensing) $f_\star (<r_{200c})=(6.8\pm 1.7)\times10^{-3}$ with DES, which is compatible with other recent measurements from an independent data set (\citealt{bahcall}).
\item On cosmological scales the ratio of baryon to total matter densities is $\Omega_b / \Omega_m \approx 16\%$ (e.g. \citealt{planck15}). In the cluster core we find  that the ratio of stellar mass to total matter is $\sim 0.7\%$. This means that if the cluster distribution is representative, then only 4\% of the baryons are locked into stars (compatible within $2\sigma$ with \citealt{fukugita}).
\item The stellar mass fraction profiles we derive from DES and CLASH are compatible within 1$\sigma$, provided that the same parametrization is used for the total matter halo profile.
\item \emph{HST} clusters can be used to test and calibrate stellar mass estimates. In future works we plan to test the stellar mass to light ratio derived from DES to that from \emph{HST}, as this should be nominally invariant under changes in galaxy apertures (that are different between the two data sets used in this work), and therefore better matched than $M_\star$ between the surveys. In addition, an extensive spectroscopic campaign carried out with the Very Large Telescope (VLT) (CLASH--VLT, \citealt{rosati}) is currently providing thousands of spectra for 14 CLASH clusters. In the future, we plan to use this survey to further test our technique: this data set will include spectra up to very large cluster radii, making CLASH--VLT ideal to test stellar masses on the large scales explored by DES. CLASH--VLT analysis of stellar masses for the cluster RXJ2248 are in course of preparation (Annunziatella et al., in preparation), and stellar mass density profiles have been studied recently for other clusters (\citealt{annunziatella14}, \citeyear{annunziatella15}).
\item In the future, we plan to apply the same techniques to a sample of $100\,000$ stacked DES clusters and deduce important information about galaxy evolution by looking at the relation between stellar and total mass, as a function of radii and redshift, at the Stellar Mass Function and at the stellar mass environment dependence. The stacking of the whole DES sample would allow us to estimate $f_\star$ with an error significantly smaller than that given here. 
\end{enumerate}

\section*{Acknowledgments}
A. Palmese acknowledges the UCL PhD studentship.
O. Lahav acknowledges support from a European Research Council Advanced Grant FP7/291329, which also supported M. Banerji and S. Jouvel.\\
D. Gruen and S. Seitz were supported by SFB-Transregio 33 ``The Dark Universe'' by the Deutsche Forschungsgemeinschaft (DFG) and the DFG cluster of excellence `Origin and Structure of the Universe'. D. Gruen was also supported by NASA through the Einstein Fellowship Program, grant PF5-160138.\\
T. Jeltema acknowledges support from the DOE grant DE-SC0013541.\\
This work has benefitted by data taken by the CLASH collaboration. \\
A. Palmese and O. Lahav acknowledge N. Bahcall and M. Milgrom for stimulating discussions about this work.\\
Funding for the DES Projects has been provided by the U.S. Department of Energy, the U.S. National Science Foundation, the Ministry of Science and Education of Spain, 
the Science and Technology Facilities Council of the United Kingdom, the Higher Education Funding Council for England, the National Center for Supercomputing 
Applications at the University of Illinois at Urbana-Champaign, the Kavli Institute of Cosmological Physics at the University of Chicago, 
the Center for Cosmology and Astro-Particle Physics at the Ohio State University,
the Mitchell Institute for Fundamental Physics and Astronomy at Texas A\&M University, Financiadora de Estudos e Projetos, 
Funda{\c c}{\~a}o Carlos Chagas Filho de Amparo {\`a} Pesquisa do Estado do Rio de Janeiro, Conselho Nacional de Desenvolvimento Cient{\'i}fico e Tecnol{\'o}gico and 
the Minist{\'e}rio da Ci{\^e}ncia, Tecnologia e Inova{\c c}{\~a}o, the Deutsche Forschungsgemeinschaft and the Collaborating Institutions in the Dark Energy Survey. 

The Collaborating Institutions are Argonne National Laboratory, the University of California at Santa Cruz, the University of Cambridge, Centro de Investigaciones Energ{\'e}ticas, 
Medioambientales y Tecnol{\'o}gicas-Madrid, the University of Chicago, University College London, the DES-Brazil Consortium, the University of Edinburgh, 
the Eidgen{\"o}ssische Technische Hochschule (ETH) Z{\"u}rich, 
Fermi National Accelerator Laboratory, the University of Illinois at Urbana-Champaign, the Institut de Ci{\`e}ncies de l'Espai (IEEC/CSIC), 
the Institut de F{\'i}sica d'Altes Energies, Lawrence Berkeley National Laboratory, the Ludwig-Maximilians Universit{\"a}t M{\"u}nchen and the associated Excellence Cluster Universe, 
the University of Michigan, the National Optical Astronomy Observatory, the University of Nottingham, The Ohio State University, the University of Pennsylvania, the University of Portsmouth, 
SLAC National Accelerator Laboratory, Stanford University, the University of Sussex, and Texas A\&M University.

The DES data management system is supported by the National Science Foundation under Grant Number AST-1138766.
The DES participants from Spanish institutions are partially supported by MINECO under grants AYA2012-39559, ESP2013-48274, FPA2013-47986, and Centro de Excelencia Severo Ochoa SEV-2012-0234.
Research leading to these results has received funding from the European Research Council under the European Unions Seventh Framework Programme (FP7/2007-2013) including ERC grant agreements 
 240672, 291329, and 306478.

\bibliographystyle{mn2e}
\bibliography{bib}

\begin{thebibliography}{55}
\expandafter\ifx\csname natexlab\endcsname\relax\def\natexlab#1{#1}\fi

\bibitem[{{Abell}, {Corwin} \& {Olowin}(1989){Abell}, {Corwin}, \&
  {Olowin}}]{abell}
{Abell} G.~O., {Corwin}, Jr. H.~G., {Olowin} R.~P., 1989, \apjs, 70, 1

\bibitem[{{Acquaviva}, {Raichoor} \& {Gawiser}(2015){Acquaviva}, {Raichoor}, \&
  {Gawiser}}]{acquaviva}
{Acquaviva} V., {Raichoor} A., {Gawiser} E., 2015, \apj, 804, 8

\bibitem[{{Andreon}(2015)}]{andreon}
{Andreon} S., 2015, \aap, 575, A108

\bibitem[{{Annunziatella} {et~al}\mbox{.}(2014){Annunziatella}, {Biviano},
  {Mercurio}, {Nonino}, {Rosati}, {Balestra}, {Presotto}, {Girardi}, {Gobat},
  {Grillo}, {Kelson}, {Medezinski}, {Postman}, {Scodeggio}, {Brescia},
  {Demarco}, {Fritz}, {Koekemoer}, {Lemze}, {Lombardi}, {Sartoris}, {Umetsu},
  {Vanzella}, {Bradley}, {Coe}, {Donahue}, {Infante}, {Kuchner}, {Maier},
  {Reg{\H o}s}, {Verdugo}, \& {Ziegler}}]{annunziatella14}
{Annunziatella} M. {et~al.}, 2014, \aap, 571, A80

\bibitem[{{Annunziatella} {et~al}\mbox{.}(2016){Annunziatella}, {Mercurio},
  {Biviano}, {Girardi}, {Nonino}, {Balestra}, {Rosati}, {Bartosch Caminha},
  {Brescia}, {Gobat}, {Grillo}, {Lombardi}, {Sartoris}, {De Lucia}, {Demarco},
  {Frye}, {Fritz}, {Moustakas}, {Scodeggio}, {Kuchner}, {Maier}, \&
  {Ziegler}}]{annunziatella15}
{Annunziatella} M. {et~al.}, 2016, \aap, 585, A160

\bibitem[{{Arnouts} {et~al}\mbox{.}(1999){Arnouts}, {Cristiani}, {Moscardini},
  {Matarrese}, {Lucchin}, {Fontana}, \& {Giallongo}}]{arnouts}
{Arnouts} S., {Cristiani} S., {Moscardini} L., {Matarrese} S., {Lucchin} F.,
  {Fontana} A., {Giallongo} E., 1999, \mnras, 310, 540

\bibitem[{{Bahcall} \& {Kulier}(2014)}]{bahcall}
{Bahcall} N.~A., {Kulier} A., 2014, \mnras, 439, 2505

\bibitem[{{Banerji} {et~al}\mbox{.}(2013){Banerji}, {Glazebrook}, {Blake},
  {Brough}, {Colless}, {Contreras}, {Couch}, {Croton}, {Croom}, {Davis},
  {Drinkwater}, \& {Forster}}]{banerji}
{Banerji} M. {et~al.}, 2013, \mnras, 431, 2209

\bibitem[{{Ben{\'{\i}}tez} {et~al}\mbox{.}(2004){Ben{\'{\i}}tez}, {Ford},
  {Bouwens}, {Menanteau}, {Blakeslee}, {Gronwall}, {Illingworth}, {Meurer},
  {Broadhurst}, {Clampin}, {Franx}, {Hartig}, {Magee}, {Sirianni}, {Ardila},
  {Bartko}, {Brown}, {Burrows}, {Cheng}, {Cross}, {Feldman}, {Golimowski},
  {Infante}, {Kimble}, {Krist}, {Lesser}, {Levay}, {Martel}, {Miley},
  {Postman}, {Rosati}, {Sparks}, {Tran}, {Tsvetanov}, {White}, \&
  {Zheng}}]{benitez}
{Ben{\'{\i}}tez} N. {et~al.}, 2004, \apjs, 150, 1

\bibitem[{{Bertin} \& {Arnouts}(1996)}]{sextractor}
{Bertin} E., {Arnouts} S., 1996, Astronomy and Astrophysics Supplement, 117,
  393

\bibitem[{{B{\"o}hringer} {et~al}\mbox{.}(2004){B{\"o}hringer}, {Schuecker},
  {Guzzo}, {Collins}, {Voges}, {Cruddace}, {Ortiz-Gil}, {Chincarini}, {De
  Grandi}, {Edge}, {MacGillivray}, {Neumann}, {Schindler}, \& {Shaver}}]{bohr}
{B{\"o}hringer} H. {et~al.}, 2004, \aap, 425, 367

\bibitem[{{Bonnett} {et~al}\mbox{.}(2015){Bonnett}, {Troxel}, {Hartley},
  {Amara}, {Leistedt}, {Becker}, {Bernstein}, {Bridle}, {Bruderer}, {Busha},
  {Carrasco Kind}, {Childress}, {Castander}, {Chang}, {Crocce}, {Davis},
  {Eifler}, {Frieman}, {Gangkofner}, {Gaztanaga}, {Glazebrook}, {Gruen},
  {Kacprzak}, {King}, {Kwan}, {Lahav}, {Lewis}, {Lidman}, {Lin}, {MacCrann},
  {Miquel}, {O'Neill}, {Palmese}, {Peiris}, {Refregier}, {Rozo}, {Rykoff},
  {Sadeh}, {S{\'a}nchez}, {Sheldon}, {Uddin}, {Wechsler}, {Zuntz}, {Abbott},
  {Abdalla}, {Allam}, {Armstrong}, {Banerji}, {Bauer}, {Benoit-L{\'e}vy},
  {Bertin}, {Brooks}, {Buckley-Geer}, {Burke}, {Capozzi}, {Carnero Rosell},
  {Carretero}, {Cunha}, {D'Andrea}, {da Costa}, {DePoy}, {Desai}, {Diehl},
  {Dietrich}, {Doel}, {Fausti Neto}, {Fernandez}, {Flaugher}, {Fosalba},
  {Gerdes}, {Gruendl}, {Honscheid}, {Jain}, {James}, {Jarvis}, {Kim}, {Kuehn},
  {Kuropatkin}, {Li}, {Lima}, {Maia}, {March}, {Marshall}, {Martini},
  {Melchior}, {Miller}, {Neilsen}, {Nichol}, {Nord}, {Ogando}, {Plazas},
  {Reil}, {Romer}, {Roodman}, {Sako}, {Sanchez}, {Santiago}, {Smith},
  {Soares-Santos}, {Sobreira}, {Suchyta}, {Swanson}, {Tarle}, {Thaler},
  {Thomas}, {Vikram}, \& {Walker}}]{bonnett}
{Bonnett} C. {et~al.}, 2015, preprint (arXiv:astro-ph/1507.05909)

\bibitem[{{Bouy} {et~al}\mbox{.}(2013){Bouy}, {Bertin}, {Moraux}, {Cuillandre},
  {Bouvier}, {Barrado}, {Solano}, \& {Bayo}}]{spreadmodel}
{Bouy} H., {Bertin} E., {Moraux} E., {Cuillandre} J.-C., {Bouvier} J.,
  {Barrado} D., {Solano} E., {Bayo} A., 2013, \aap, 554, A101

\bibitem[{Bruzual \& Charlot(2003)}]{bc}
Bruzual G., Charlot S., 2003, \mnras, 344, 1000

\bibitem[{Calzetti {et~al}\mbox{.}(2000)Calzetti, Armus, Bohlin, Kinney,
  Koornneef, {et~al.}}]{calzetti}
Calzetti D., Armus L., Bohlin R.~C., Kinney A.~L., Koornneef J., {et~al.},
  2000, Astrophys.J., 533, 682

\bibitem[{Chabrier(2003)}]{ch}
Chabrier G., 2003, Publ.Astron.Soc.Pac., 115, 763

\bibitem[{{Clowe} {et~al}\mbox{.}(1998){Clowe}, {Luppino}, {Kaiser}, {Henry},
  \& {Gioia}}]{clowe}
{Clowe} D., {Luppino} G.~A., {Kaiser} N., {Henry} J.~P., {Gioia} I.~M., 1998,
  \apj, 497, L61

\bibitem[{{Conroy}(2013)}]{conroy}
{Conroy} C., 2013, Annual Review of Astronomy and Astrophysics, 51, 393

\bibitem[{{Crocce} {et~al}\mbox{.}(2016){Crocce}, {Carretero}, {Bauer}, {Ross},
  {Sevilla-Noarbe}, {Giannantonio}, {Sobreira}, {Sanchez}, {Gaztanaga}, {Kind},
  {S{\'a}nchez}, {Bonnett}, {Benoit-L{\'e}vy}, {Brunner}, {Rosell}, {Cawthon},
  {Fosalba}, {Hartley}, {Kim}, {Leistedt}, {Miquel}, {Peiris}, {Percival},
  {Rosenfeld}, {Rykoff}, {S{\'a}nchez}, {Abbott}, {Abdalla}, {Allam},
  {Banerji}, {Bernstein}, {Bertin}, {Brooks}, {Buckley-Geer}, {Burke},
  {Capozzi}, {Castander}, {Cunha}, {D'Andrea}, {da Costa}, {Desai}, {Diehl},
  {Eifler}, {Evrard}, {Neto}, {Fernandez}, {Finley}, {Flaugher}, {Frieman},
  {Gerdes}, {Gruen}, {Gruendl}, {Gutierrez}, {Honscheid}, {James}, {Kuehn},
  {Kuropatkin}, {Lahav}, {Li}, {Lima}, {Maia}, {March}, {Marshall}, {Martini},
  {Melchior}, {Miller}, {Neilsen}, {Nichol}, {Nord}, {Ogando}, {Plazas},
  {Romer}, {Sako}, {Santiago}, {Schubnell}, {Smith}, {Soares-Santos},
  {Suchyta}, {Swanson}, {Tarle}, {Thaler}, {Thomas}, {Vikram}, {Walker},
  {Wechsler}, {Weller}, {Zuntz}, \& {DES Collaboration}}]{crocce}
{Crocce} M. {et~al.}, 2016, \mnras, 455, 4301

\bibitem[{{de Blok}(2010)}]{cusp}
{de Blok} W.~J.~G., 2010, Advances in Astronomy, 2010, 789293

\bibitem[{{Desai} {et~al}\mbox{.}(2012){Desai}, {Armstrong}, {Mohr}, {Semler},
  {Liu}, {Bertin}, {Allam}, {Barkhouse}, {Bazin}, {Buckley-Geer}, {Cooper},
  {Hansen}, {High}, {Lin}, {Lin}, {Ngeow}, {Rest}, {Song}, {Tucker}, \&
  {Zenteno}}]{desai}
{Desai} S. {et~al.}, 2012, \apj, 757, 83

\bibitem[{{Diehl} {et~al}\mbox{.}(2014){Diehl}, {Abbott}, {Annis}, {Armstrong},
  {Baruah}, {Bermeo}, {Bernstein}, {Beynon}, {Bruderer}, {Buckley-Geer},
  {Campbell}, {Capozzi}, {Carter}, {Casas}, {Clerkin}, {Covarrubias}, {Cuhna},
  {D'Andrea}, {da Costa}, {Das}, {DePoy}, {Dietrich}, {Drlica-Wagner},
  {Elliott}, {Eifler}, {Estrada}, {Etherington}, {Flaugher}, {Frieman}, {Fausti
  Neto}, {Gelman}, {Gerdes}, {Gruen}, {Gruendl}, {Hao}, {Head}, {Helsby},
  {Hoffman}, {Honscheid}, {James}, {Johnson}, {Kacprzac}, {Katsaros},
  {Kennedy}, {Kent}, {Kessler}, {Kim}, {Krause}, {Kron}, {Kuhlmann}, {Kunder},
  {Li}, {Lin}, {Maccrann}, {March}, {Marshall}, {Neilsen}, {Nugent}, {Martini},
  {Melchior}, {Menanteau}, {Nichol}, {Nord}, {Ogando}, {Old}, {Papadopoulos},
  {Patton}, {Petravick}, {Plazas}, {Poulton}, {Pujol}, {Reil}, {Rigby},
  {Romer}, {Roodman}, {Rooney}, {Sanchez Alvaro}, {Serrano}, {Sheldon},
  {Smith}, {Smith}, {Soares-Santos}, {Soumagnac}, {Spinka}, {Suchyta},
  {Tucker}, {Walker}, {Wester}, {Wiesner}, {Wilcox}, {Williams}, {Yanny}, \&
  {Zhang}}]{y1}
{Diehl} H.~T. {et~al.}, 2014, in Society of Photo-Optical Instrumentation
  Engineers (SPIE) Conference Series, Vol. 9149, Society of Photo-Optical
  Instrumentation Engineers (SPIE) Conference Series, p. 91490V

\bibitem[{{Dietrich} {et~al}\mbox{.}(2012){Dietrich}, {B{\"o}hnert},
  {Lombardi}, {Hilbert}, \& {Hartlap}}]{dietrich}
{Dietrich} J.~P., {B{\"o}hnert} A., {Lombardi} M., {Hilbert} S., {Hartlap} J.,
  2012, \mnras, 419, 3547

\bibitem[{{Eisenstein} {et~al}\mbox{.}(2001){Eisenstein}, {Annis}, {Gunn},
  {Szalay}, {Connolly}, {Nichol}, {Bahcall}, {Bernardi}, {Burles}, {Castander},
  {Fukugita}, {Hogg}, {Ivezi{\'c}}, {Knapp}, {Lupton}, {Narayanan}, {Postman},
  {Reichart}, {Richmond}, {Schneider}, {Schlegel}, {Strauss}, {SubbaRao},
  {Tucker}, {Vanden Berk}, {Vogeley}, {Weinberg}, \& {Yanny}}]{eisenstein}
{Eisenstein} D.~J. {et~al.}, 2001, \aj, 122, 2267

\bibitem[{{Flaugher} {et~al}\mbox{.}(2015){Flaugher}, {Diehl}, {Honscheid},
  {Abbott}, {Alvarez}, {Angstadt}, {Annis}, {Antonik}, {Ballester}, {Beaufore},
  {Bernstein}, {Bernstein}, {Bigelow}, {Bonati}, {Boprie}, {Brooks},
  {Buckley-Geer}, {Campa}, {Cardiel-Sas}, {Castander}, {Castilla}, {Cease},
  {Cela-Ruiz}, {Chappa}, {Chi}, {Cooper}, {da Costa}, {Dede}, {Derylo},
  {DePoy}, {de Vicente}, {Doel}, {Drlica-Wagner}, {Eiting}, {Elliott}, {Emes},
  {Estrada}, {Fausti Neto}, {Finley}, {Flores}, {Frieman}, {Gerdes},
  {Gladders}, {Gregory}, {Gutierrez}, {Hao}, {Holland}, {Holm}, {Huffman},
  {Jackson}, {James}, {Jonas}, {Karcher}, {Karliner}, {Kent}, {Kessler},
  {Kozlovsky}, {Kron}, {Kubik}, {Kuehn}, {Kuhlmann}, {Kuk}, {Lahav}, {Lathrop},
  {Lee}, {Levi}, {Lewis}, {Li}, {Mandrichenko}, {Marshall}, {Martinez},
  {Merritt}, {Miquel}, {Mu{\~n}oz}, {Neilsen}, {Nichol}, {Nord}, {Ogando},
  {Olsen}, {Palaio}, {Patton}, {Peoples}, {Plazas}, {Rauch}, {Reil}, {Rheault},
  {Roe}, {Rogers}, {Roodman}, {Sanchez}, {Scarpine}, {Schindler}, {Schmidt},
  {Schmitt}, {Schubnell}, {Schultz}, {Schurter}, {Scott}, {Serrano}, {Shaw},
  {Smith}, {Soares-Santos}, {Stefanik}, {Stuermer}, {Suchyta}, {Sypniewski},
  {Tarle}, {Thaler}, {Tighe}, {Tran}, {Tucker}, {Walker}, {Wang}, {Watson},
  {Weaverdyck}, {Wester}, {Woods}, {Yanny}, \& {The DES
  Collaboration}}]{flaugher}
{Flaugher} B. {et~al.}, 2015, \aj, 150, 150

\bibitem[{{Fukugita} \& {Peebles}(2004)}]{fukugita}
{Fukugita} M., {Peebles} P.~J.~E., 2004, \apj, 616, 643

\bibitem[{{G{\'o}mez} {et~al}\mbox{.}(2012){G{\'o}mez}, {Valkonen}, {Romer},
  {Lloyd-Davies}, {Verdugo}, {Cantalupo}, {Daub}, {Goldstein}, {Kuo}, {Lange},
  {Lueker}, {Holzapfel}, {Peterson}, {Ruhl}, {Runyan}, {Reichardt}, \&
  {Sabirli}}]{gomez}
{G{\'o}mez} P.~L. {et~al.}, 2012, \aj, 144, 79

\bibitem[{{G{\'o}rski} {et~al}\mbox{.}(2005){G{\'o}rski}, {Hivon}, {Banday},
  {Wandelt}, {Hansen}, {Reinecke}, \& {Bartelmann}}]{healpix}
{G{\'o}rski} K.~M., {Hivon} E., {Banday} A.~J., {Wandelt} B.~D., {Hansen}
  F.~K., {Reinecke} M., {Bartelmann} M., 2005, \apj, 622, 759

\bibitem[{{Gruen} {et~al}\mbox{.}(2013){Gruen}, {Brimioulle}, {Seitz}, {Lee},
  {Young}, {Koppenhoefer}, {Eichner}, {Riffeser}, {Vikram}, {Weidinger}, \&
  {Zenteno}}]{gruen}
{Gruen} D. {et~al.}, 2013, MNRAS, 432, 1455

\bibitem[{{Ilbert} {et~al}\mbox{.}(2006){Ilbert}, {Arnouts}, {McCracken},
  {Bolzonella}, {Bertin}, {Le F{\`e}vre}, {Mellier}, {Zamorani}, {Pell{\`o}},
  {Iovino}, {Tresse}, {Le Brun}, {Bottini}, {Garilli}, {Maccagni}, \&
  {Picat}}]{ilbertlephare}
{Ilbert} O. {et~al.}, 2006, \aap, 457, 841

\bibitem[{{Ilbert} {et~al}\mbox{.}(2009){Ilbert}, {Capak}, {Salvato}, {Aussel},
  {McCracken}, {Sanders}, {Scoville}, {Kartaltepe}, {Arnouts}, {Le Floc'h},
  {Mobasher}, {Taniguchi}, {Lamareille}, \& {Leauthaud}}]{cosmos}
{Ilbert} O. {et~al.}, 2009, \apj, 690, 1236

\bibitem[{{Ilbert} {et~al}\mbox{.}(2010){Ilbert}, {Salvato}, {Le Floc'h},
  {Aussel}, {Capak}, {McCracken}, {Mobasher}, {Kartaltepe}, {Scoville},
  {Sanders}, {Arnouts}, {Bundy}, {Cassata}, \& {Kneib}}]{ilbert2010}
{Ilbert} O. {et~al.}, 2010, \apj, 709, 644

\bibitem[{{Jouvel} {et~al}\mbox{.}(2014){Jouvel}, {Host}, {Lahav}, {Seitz},
  {Molino}, {Coe}, {Postman}, {Moustakas}, {Ben{\`i}tez}, {Rosati}, {Balestra},
  {Grillo}, {Bradley}, {Fritz}, {Kelson}, \& {Koekemoer}}]{jouvel}
{Jouvel} S. {et~al.}, 2014, \aap, 562, A86

\bibitem[{{Koekemoer} {et~al}\mbox{.}(2011){Koekemoer}, {Faber}, {Ferguson},
  {Grogin}, {Kocevski}, {Koo}, {Lai}, {Lotz}, {Lucas}, {McGrath}, {Ogaz},
  {Rajan}, {Riess}, {Rodney}, {Strolger}, {Casertano}, {Castellano}, {Dahlen},
  {Dickinson}, {Dolch}, {Fontana}, {Giavalisco}, {Grazian}, {Guo}, {Hathi},
  {Huang}, {van der Wel}, {Yan}, {Acquaviva}, {Alexander}, {Almaini}, {Ashby},
  {Barden}, {Bell}, {Bournaud}, {Brown}, {Caputi}, {Cassata}, {Challis},
  {Chary}, {Cheung}, {Cirasuolo}, {Conselice}, {Roshan Cooray}, {Croton},
  {Daddi}, {Dav{\'e}}, {de Mello}, {de Ravel}, {Dekel}, {Donley}, {Dunlop},
  {Dutton}, {Elbaz}, {Fazio}, {Filippenko}, {Finkelstein}, {Frazer}, {Gardner},
  {Garnavich}, {Gawiser}, {Gruetzbauch}, {Hartley}, {H{\"a}ussler},
  {Herrington}, {Hopkins}, {Huang}, {Jha}, {Johnson}, {Kartaltepe},
  {Khostovan}, {Kirshner}, {Lani}, {Lee}, {Li}, {Madau}, {McCarthy},
  {McIntosh}, {McLure}, {McPartland}, {Mobasher}, {Moreira}, {Mortlock},
  {Moustakas}, {Mozena}, {Nandra}, {Newman}, {Nielsen}, {Niemi}, {Noeske},
  {Papovich}, {Pentericci}, {Pope}, {Primack}, {Ravindranath}, {Reddy},
  {Renzini}, {Rix}, {Robaina}, {Rosario}, {Rosati}, {Salimbeni}, {Scarlata},
  {Siana}, {Simard}, {Smidt}, {Snyder}, {Somerville}, {Spinrad}, {Straughn},
  {Telford}, {Teplitz}, {Trump}, {Vargas}, {Villforth}, {Wagner}, {Wandro},
  {Wechsler}, {Weiner}, {Wiklind}, {Wild}, {Wilson}, {Wuyts}, \&
  {Yun}}]{koekemoer}
{Koekemoer} A.~M. {et~al.}, 2011, \apjs, 197, 36

\bibitem[{{Maughan} {et~al}\mbox{.}(2008){Maughan}, {Jones}, {Forman}, \& {Van
  Speybroeck}}]{maughan08}
{Maughan} B.~J., {Jones} C., {Forman} W., {Van Speybroeck} L., 2008, \apjs,
  174, 117

\bibitem[{{Melchior} {et~al}\mbox{.}(2015){Melchior}, {Suchyta}, {Huff},
  {Hirsch}, {Kacprzak}, {Rykoff}, {Gruen}, {Armstrong}, {Bacon}, {Bechtol},
  {Bernstein}, {Bridle}, {Clampitt}, {Honscheid}, {Jain}, \&
  {Jouvel}}]{melchior}
{Melchior} P. {et~al.}, 2015, \mnras, 449, 2219

\bibitem[{{Mohr} {et~al}\mbox{.}(2012){Mohr}, {Armstrong}, {Bertin}, {Daues},
  {Desai}, {Gower}, {Gruendl}, {Hanlon}, {Kuropatkin}, {Lin}, {Marriner},
  {Petravic}, {Sevilla}, {Swanson}, {Tomashek}, {Tucker}, \&
  {Yanny}}]{dataproc}
{Mohr} J.~J. {et~al.}, 2012, Society of Photo-Optical Instrumentation Engineers
  (SPIE) Conference Series, 8451, 0

\bibitem[{{Monna} {et~al}\mbox{.}(2014){Monna}, {Seitz}, {Greisel}, {Eichner},
  {Drory}, {Postman}, {Zitrin}, {Coe}, {Halkola}, {Suyu}, {Grillo}, {Rosati},
  {Lemze}, {Balestra}, {Snigula}, \& {Bradley}}]{monna}
{Monna} A. {et~al.}, 2014, \mnras, 438, 1417

\bibitem[{{Navarro}, {Frenk} \& {White}(1996){Navarro}, {Frenk}, \&
  {White}}]{nfw}
{Navarro} J.~F., {Frenk} C.~S., {White} S.~D.~M., 1996, \apj, 462, 563

\bibitem[{{Oaxaca Wright} \& {Brainerd}(2000)}]{nfwfit}
{Oaxaca Wright} C., {Brainerd} T.~G., 2000, \aj, 534, 34

\bibitem[{{Planck Collaboration} {et~al}\mbox{.}(2015){Planck Collaboration},
  {Ade}, {Aghanim}, {Arnaud}, {Ashdown}, {Aumont}, {Baccigalupi}, {Banday},
  {Barreiro}, {Bartlett}, \& et~al.}]{planck15}
{Planck Collaboration} {et~al.}, 2015, XIII, preprint
  (arXiv:astro-ph/1502.01589)

\bibitem[{{Postman} {et~al}\mbox{.}(2012){Postman}, {Coe}, {Ben{\'{\i}}tez},
  {Bradley}, {Broadhurst}, {Donahue}, {Ford}, {Graur}, {Graves}, {Jouvel},
  {Koekemoer}, {Lemze}, {Medezinski}, {Molino}, {Moustakas}, \&
  {Ogaz}}]{postman}
{Postman} M. {et~al.}, 2012, \apjs, 199, 25

\bibitem[{{Rosati} {et~al}\mbox{.}(2014){Rosati}, {Balestra}, {Grillo},
  {Mercurio}, {Nonino}, {Biviano}, {Girardi}, {Vanzella}, \& {Clash-VLT
  Team}}]{rosati}
{Rosati} P. {et~al.}, 2014, The Messenger, 158, 48

\bibitem[{{Rozo} {et~al}\mbox{.}(2015){Rozo}, {Rykoff}, {Becker}, {Reddick}, \&
  {Wechsler}}]{rozo}
{Rozo} E., {Rykoff} E.~S., {Becker} M., {Reddick} R.~M., {Wechsler} R.~H.,
  2015, \mnras, 453, 38

\bibitem[{{Rykoff} {et~al}\mbox{.}(2014){Rykoff}, {Rozo}, {Busha}, {Cunha},
  {Finoguenov}, {Evrard}, {Hao}, {Koester}, {Leauthaud}, {Nord}, {Pierre},
  {Reddick}, {Sadibekova}, {Sheldon}, \& {Wechsler}}]{redpaper}
{Rykoff} E.~S. {et~al.}, 2014, \apj, 785, 104

\bibitem[{{Rykoff} {et~al}\mbox{.}(2016){Rykoff}, {Rozo}, {Hollowood},
  {Bermeo-Hernandez}, {Jeltema}, {Mayers}, {Romer}, {Rooney}, {Saro}, {Vergara
  Cervantes}, {Wilcox}, {Abbott}, {Abdalla}, {Allam}, {Annis},
  {Benoit-L{\'e}vy}, {Bernstein}, {Bertin}, {Brooks}, {Burke}, {Capozzi},
  {Carnero Rosell}, {Carrasco Kind}, {Castander}, {Childress}, {Collins},
  {Cunha}, {D'Andrea}, {da Costa}, {Davis}, {Desai}, {Diehl}, {Dietrich},
  {Doel}, {Evrard}, {Finley}, {Flaugher}, {Fosalba}, {Frieman}, {Glazebrook},
  {Goldstein}, {Gruen}, {Gruendl}, {Gutierrez}, {Hilton}, {Honscheid}, {Hoyle},
  {James}, {Kay}, {Kuehn}, {Kuropatkin}, {Lahav}, {Lewis}, {Lidman}, {Lima},
  {Maia}, {Mann}, {Marshall}, {Martini}, {Melchior}, {Miller}, {Miquel},
  {Mohr}, {Nichol}, {Nord}, {Ogando}, {Plazas}, {Reil}, {Sahl{\'e}n},
  {Sanchez}, {Santiago}, {Scarpine}, {Schubnell}, {Sevilla-Noarbe}, {Smith},
  {Soares-Santos}, {Sobreira}, {Stott}, {Suchyta}, {Swanson}, {Tarle},
  {Thomas}, {Tucker}, {Viana}, {Vikram}, {Walker}, \& {Zhang}}]{rykoff}
{Rykoff} E.~S. {et~al.}, 2016, \apjs, 224, 1

\bibitem[{{S{\'a}nchez} {et~al}\mbox{.}(2014){S{\'a}nchez}, {Carrasco Kind},
  {Lin}, {Miquel}, {Abdalla}, {Amara}, {Banerji}, {Bonnett}, {Brunner},
  {Capozzi}, {Carnero}, {Castander}, {da Costa}, {Cunha}, {Fausti}, {Gerdes},
  {Greisel}, {Gschwend}, {Hartley}, {Jouvel}, {Lahav}, {Lima}, {Maia},
  {Mart{\'{\i}}}, {Ogando}, {Ostrovski}, {Pellegrini}, {Rau}, {Sadeh}, {Seitz},
  {Sevilla-Noarbe}, {Sypniewski}, {de Vicente}, {Abbot}, {Allam}, {Atlee},
  {Bernstein}, {Bernstein}, {Buckley-Geer}, {Burke}, {Childress}, {Davis},
  {DePoy}, {Dey}, {Desai}, {Diehl}, {Doel}, {Estrada}, {Evrard},
  {Fern{\'a}ndez}, {Finley}, {Flaugher}, {Frieman}, {Gaztanaga}, {Glazebrook},
  {Honscheid}, {Kim}, {Kuehn}, {Kuropatkin}, {Lidman}, {Makler}, {Marshall},
  {Nichol}, {Roodman}, {S{\'a}nchez}, {Santiago}, {Sako}, {Scalzo}, {Smith},
  {Swanson}, {Tarle}, {Thomas}, {Tucker}, {Uddin}, {Vald{\'e}s}, {Walker},
  {Yuan}, \& {Zuntz}}]{sanchez}
{S{\'a}nchez} C. {et~al.}, 2014, \mnras, 445, 1482

\bibitem[{{Schneider}(1996)}]{schneider1996}
{Schneider} P., 1996, \mnras, 283, 837

\bibitem[{{Sevilla} {et~al}\mbox{.}(2011){Sevilla}, {Armstrong}, {Bertin},
  {Carlson}, {Daues}, {Desai}, {Gower}, {Gruendl}, {Hanlon}, {Jarvis},
  {Kessler}, {Kuropatkin}, {Lin}, {Marriner}, {Mohr}, {Petravick}, {Sheldon},
  {Swanson}, {Tomashek}, {Tucker}, {Yang}, {Yanny}, \& {for the DES
  Collaboration}}]{sevilla}
{Sevilla} I. {et~al.}, 2011, preprint (arXiv:astro-ph/1109.6741)

\bibitem[{{Soumagnac} {et~al}\mbox{.}(2015){Soumagnac}, {Abdalla}, {Lahav},
  {Kirk}, {Sevilla}, {Bertin}, {Rowe}, {Annis}, {Busha}, {Da Costa}, {Frieman},
  {Gaztanaga}, {Jarvis}, {Lin}, {Percival}, {Santiago}, {Sabiu}, {Wechsler},
  {Wolz}, \& {Yanny}}]{soumagnac}
{Soumagnac} M.~T. {et~al.}, 2015, \mnras, 450, 666

\bibitem[{{Suchyta} {et~al}\mbox{.}(2016){Suchyta}, {Huff}, {Aleksi{\'c}},
  {Melchior}, {Jouvel}, {MacCrann}, {Ross}, {Crocce}, {Gaztanaga}, {Honscheid},
  {Leistedt}, {Peiris}, {Rykoff}, {Sheldon}, {Abbott}, {Abdalla}, {Allam},
  {Banerji}, {Benoit-L{\'e}vy}, {Bertin}, {Brooks}, {Burke}, {Rosell}, {Kind},
  {Carretero}, {Cunha}, {D'Andrea}, {da Costa}, {DePoy}, {Desai}, {Diehl},
  {Dietrich}, {Doel}, {Eifler}, {Estrada}, {Evrard}, {Flaugher}, {Fosalba},
  {Frieman}, {Gerdes}, {Gruen}, {Gruendl}, {James}, {Jarvis}, {Kuehn},
  {Kuropatkin}, {Lahav}, {Lima}, {Maia}, {March}, {Marshall}, {Miller},
  {Miquel}, {Neilsen}, {Nichol}, {Nord}, {Ogando}, {Percival}, {Reil},
  {Roodman}, {Sako}, {Sanchez}, {Scarpine}, {Sevilla-Noarbe}, {Smith},
  {Soares-Santos}, {Sobreira}, {Swanson}, {Tarle}, {Thaler}, {Thomas},
  {Vikram}, {Walker}, {Wechsler}, {Zhang}, \& {DES Collaboration}}]{suchyta}
{Suchyta} E. {et~al.}, 2016, \mnras, 457, 786

\bibitem[{{Swanson} {et~al}\mbox{.}(2008){Swanson}, {Tegmark}, {Hamilton}, \&
  {Hill}}]{mangle}
{Swanson} M.~E.~C., {Tegmark} M., {Hamilton} A.~J.~S., {Hill} J.~C., 2008,
  \mnras, 387, 1391

\bibitem[{{The Dark Energy Survey Collaboration}(2005)}]{descollaboration}
{The Dark Energy Survey Collaboration}, 2005, preprint (arXiv:astro-ph/0510346)

\bibitem[{{Umetsu} {et~al}\mbox{.}(2016){Umetsu}, {Zitrin}, {Gruen}, {Merten},
  {Donahue}, \& {Postman}}]{umetsu15}
{Umetsu} K., {Zitrin} A., {Gruen} D., {Merten} J., {Donahue} M., {Postman} M.,
  2016, \apj, 821, 116

\bibitem[{{Zhang} {et~al}\mbox{.}(2015){Zhang}, {McKay}, {Bertin}, {Jeltema},
  {Miller}, {Rykoff}, \& {Song}}]{yuanyuan}
{Zhang} Y., {McKay} T.~A., {Bertin} E., {Jeltema} T., {Miller} C.~J., {Rykoff}
  E., {Song} J., 2015, Publications of the Astronomical Society of the Pacific,
  127, 1183

\end{thebibliography}

\appendix

\section{Magnitude comparison}\label{magcomp}
\begin{figure*}\centering
\hskip -1.5cm\includegraphics[width=8.1cm]{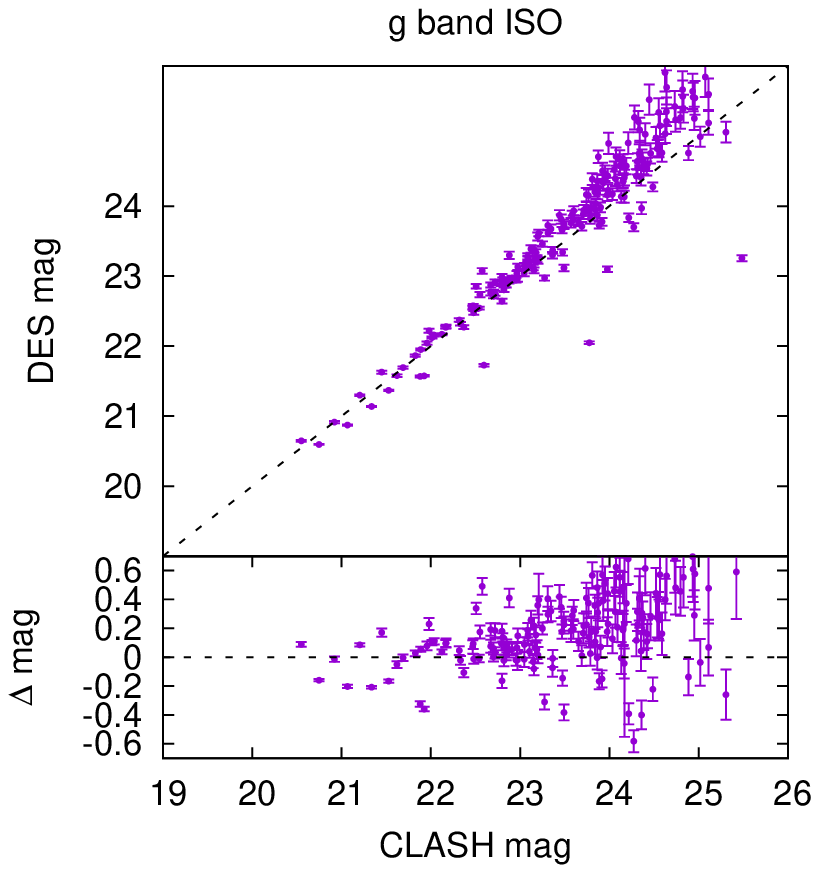}\hskip 3cm
\includegraphics[width=8.1cm]{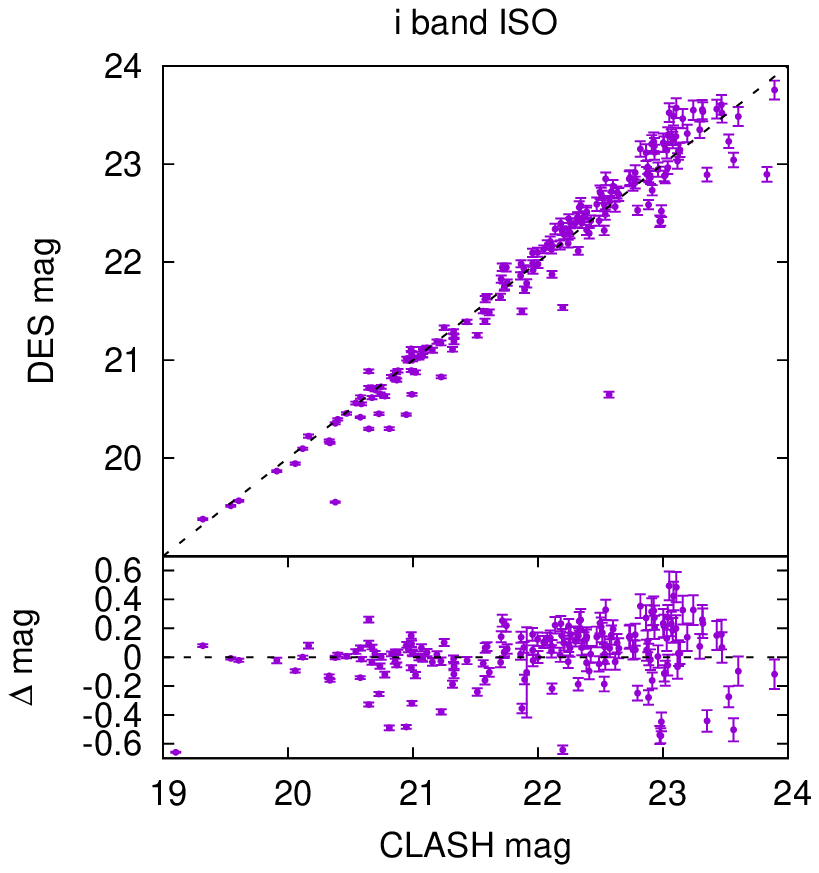}
\vskip -5.8cm\hskip -1cm
\includegraphics[width=8.1cm]{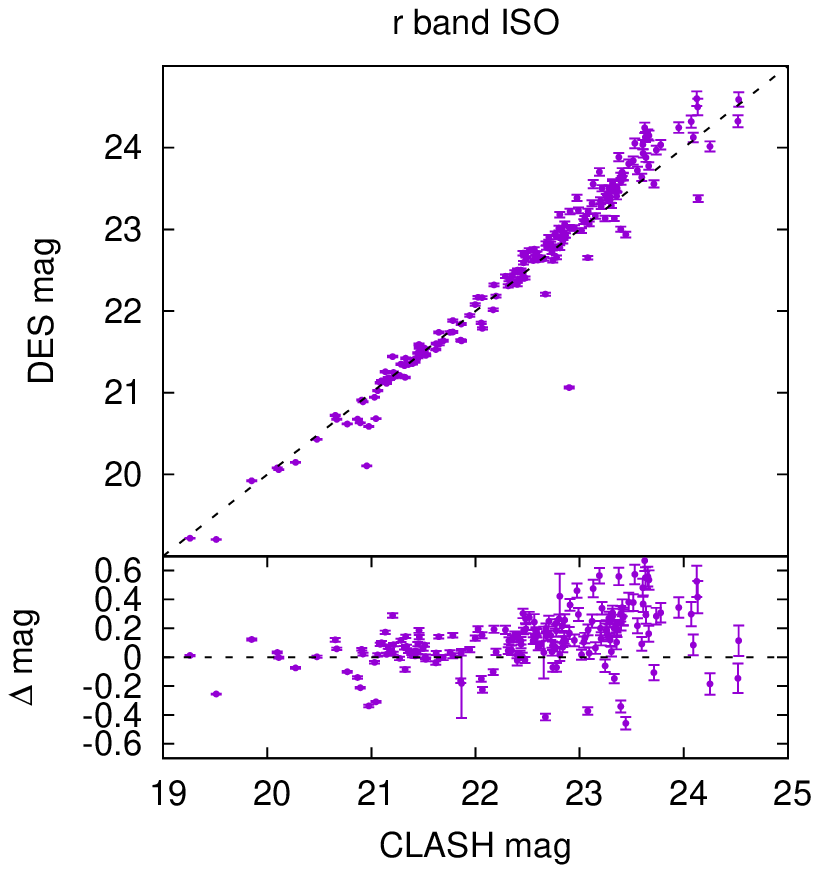}\\
\includegraphics[width=8.1cm]{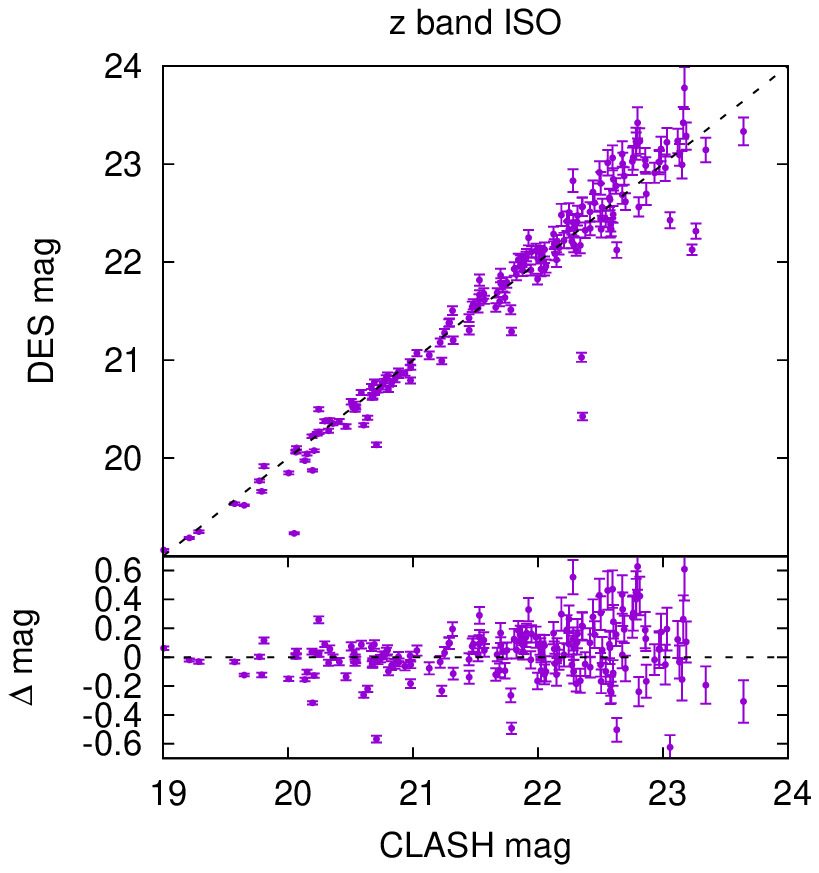}
\includegraphics[width=8.1cm]{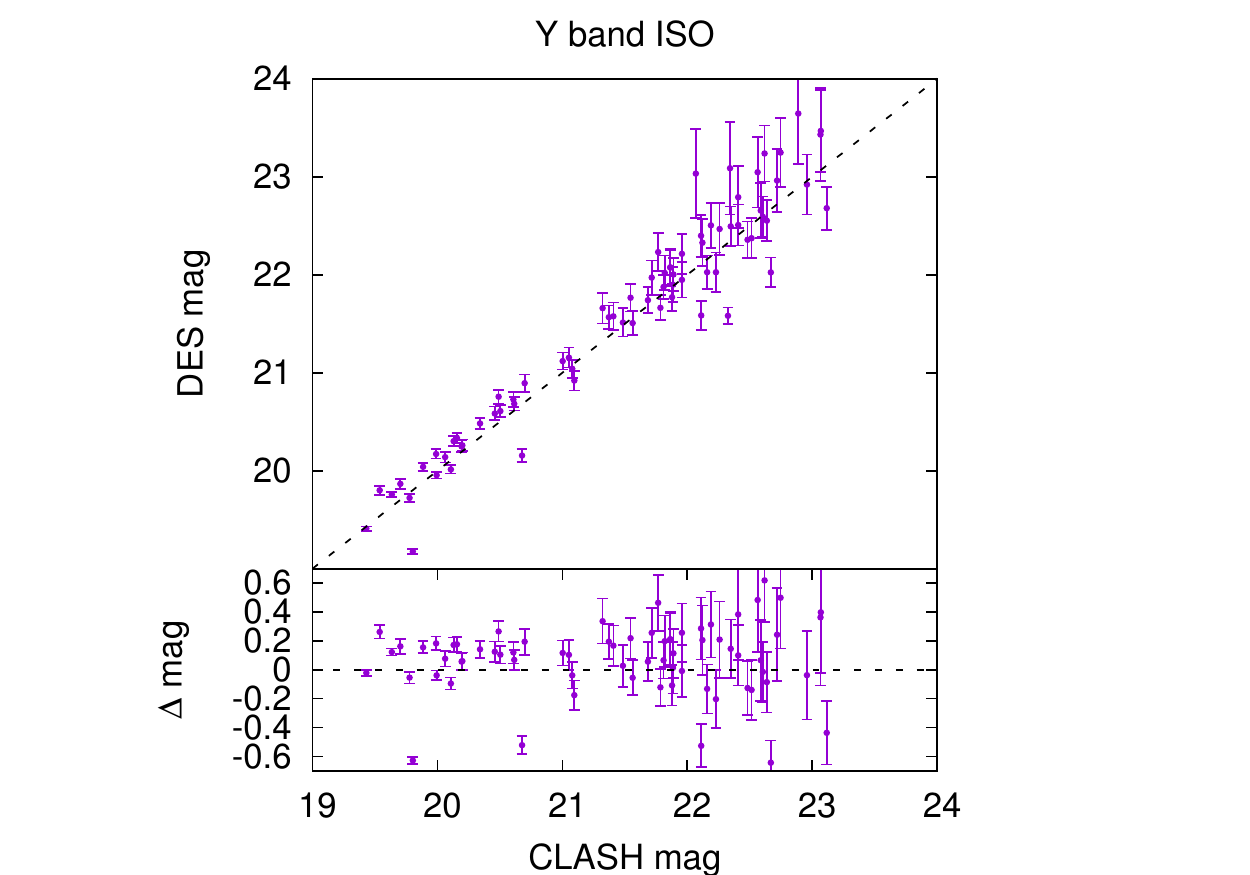}
\caption{ DES magnitues compared to CLASH magnitudes, with bottom plots of $\Delta_{\rm m} =  m_{\rm CLASH} - m_{\rm DES}$ in the $g$,$r$,$i$ and $z$ bands for the matched sources that satisfy $S/N>10$ and filtering the sources with \texttt{FLAGS}$> 3$. CLASH errorbars are not plotted for visualisation purposes, while those on the DES magnitudes represent the $1\sigma$ error. The dashed black lines represent the ideal case $m_{\rm CLASH} = m_{\rm DES}$. }\label{deltamag}
\end{figure*}
Considering only those matched sources with a signal-to-noise ratio $S/N>10$ in the DES $i$-band, excluding stars and objects with SExtractor $\texttt{FLAGS} > 3$ (in order to exclude objects with saturated pixels or corrupted data, but include objects that were initially blended) we are left with 327 sources observed in the $g$ and $r$ bands, and 331 in the $i$ and $z$ bands. The differences $\Delta_{\rm m} =  m_{\rm DES}- m_{\rm CLASH}$ are plotted in Figure \ref{deltamag}, as well as the DES magnitudes as a function of the CLASH ones for the matched sources. The magnitudes plotted are SExtractor isophotal magnitudes \texttt{MAG\_ISO} for both DES and CLASH. $\Delta_{\rm m}$ in the $griz$ bands has been corrected for the magnitude shifts due to the differences between these DES and \emph{HST} filters. The offsets have been computed using two ``extreme case'' SED templates (one elliptical, one irregular) at the cluster redshift. We have not taken into consideration the $Y$ band offset as the \emph{HST} and DES filters are too different, but we still report the comparison for completeness.

Figure \ref{deltamag} shows an offset in the DES magnitudes, especially in the $gri$ bands, which may be due to different choices of threshold or background when running SExtractor. Nevertheless the linear trend is clear, bringing to Pearson coefficients between 0.91 and 0.98 in all bands. The higher scatter that we could expect in the $Y$ band (as the corresponding CLASH filter is the F105W, which is much more spread towards the infrared than the DES $Y$ filter) is actually compensated for by higher DES photometric errors. The mean difference in magnitude $\Delta_{\rm m}$ between the two data sets is 0.13, 0.04, -0.07, -0.07 and 0.08 in the $grizY$ bands respectively.

\section{photo-$z$ consistency test}\label{test}
In this Section we test whether the fact that we assumed a photo-$z$ that was computed with a certain set of templates to then compute the stellar mass with a larger set of templates is consistent. With photometric surveys data, it is a common method to perform estimation of the photometric redshift and SED fitting in two steps, which involve fixing the redshift of a galaxy at the best fit value obtained in the first step. Although this may not be the most elegant way of solving the problem of the lack of spectroscopic information, it has been proven to lead to only a small bias in the SED fitting parameters, when compared to results given by the simultaneous estimation of redshift and stellar mass (see \citealt{acquaviva}). Moreover, here we can take advantage of the prior information that this is a cluster.

\begin{figure*}
\includegraphics[width=0.4\textwidth]{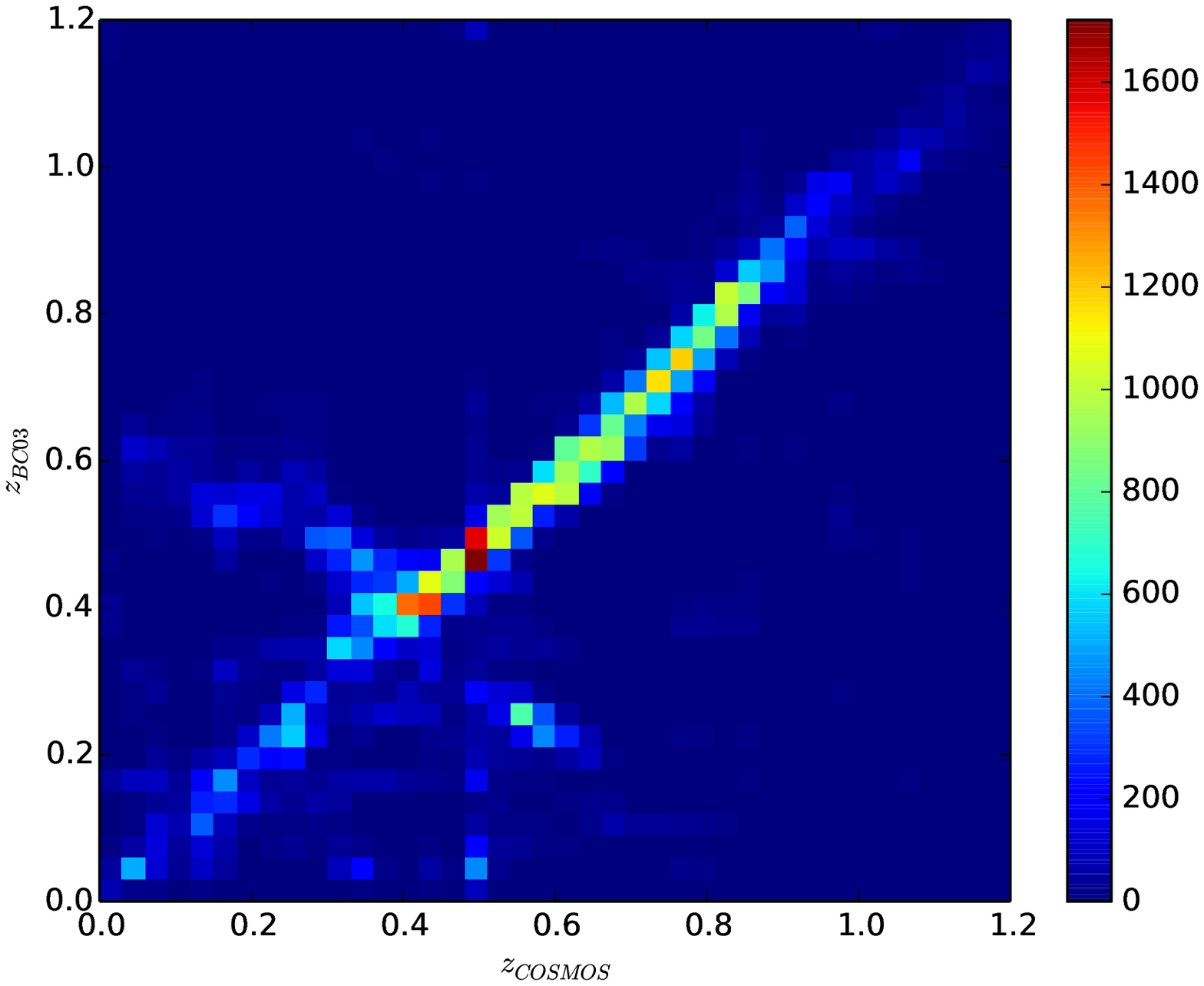}
\includegraphics[width=0.4\textwidth]{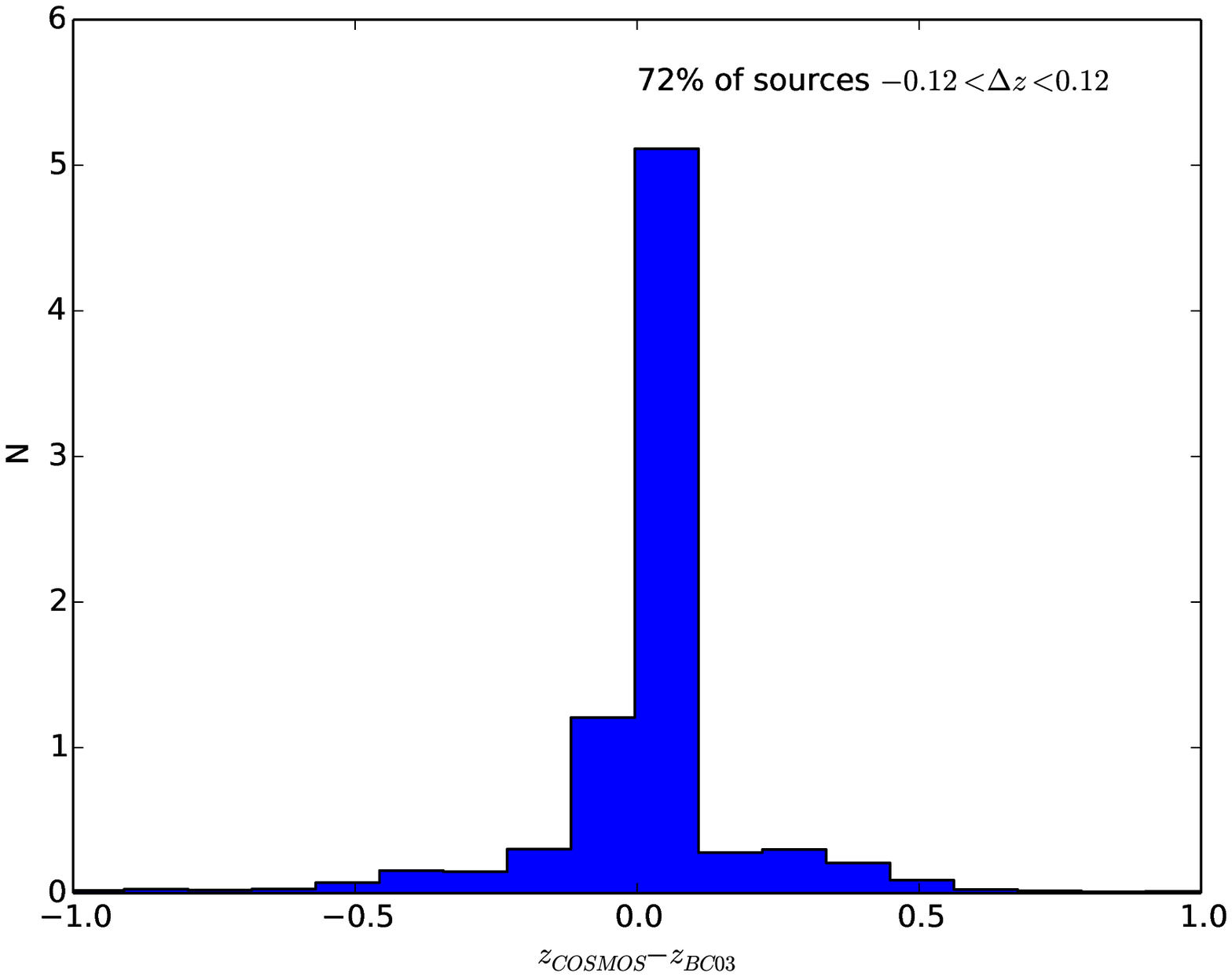}\caption{\emph{Left panel}: Comparison  of the photo-$z$'s computed using the Bruzual and Charlot 2003 templates and those using the COSMOS templates. \emph{Right panel:} Residuals of the photo-$z$'s computed using the two differnt set of templates.}\label{bc03vscosmos}
\end{figure*}

In order to do test the consistency of the photo-$z$ choice, we want to show that the resulting photo-$z$ is not drastically dependent on the choice of templates. Therefore, we compare the photo-$z$'s given by Le Phare when using the COSMOS templates and the BC03 ones. A comparison is shown in Figure \ref{bc03vscosmos} for all the galaxies in the DES field of view. We retrieve that $71\%$ of the galaxies have $|\Delta z|=|z_{BC03}-z_{COSMOS}|<0.12$.

\bsp
\newpage
\section{Affiliations}
$^{1}$Department of Physics and Astronomy, University College London, Gower Street, London, WC1E 6BT, UK\\
$^{2}$Institute of Astronomy, University of Cambidge, Madingley Road, Cambridge, CB3 0HA, UK\\
$^{3}$SLAC National Accelerator Laboratory, Menlo Park, CA 94025, USA\\
$^{4}$Kavli Institute for Particle Astrophysics \& Cosmology, P. O. Box 2450, Stanford University, Stanford, CA 94305, USA\\
$^{5}$University Observatory Munich, Scheinerstrasse 1, 81679 Munich, Germany\\
$^{6}$Max Planck Institute for Extraterrestrial Physics, Giessenbachstrasse, 85748 Garching, Germany\\
$^{7}$Department of Astrophysical Sciences, Princeton University, Princeton, NJ 08544, USA\\
$^{8}$Institut de F\'{\i}sica d'Altes Energies (IFAE), The Barcelona Institute of Science and Technology, Campus UAB, 08193 Bellaterra (Barcelona) Spain\\
$^{9}$Fermi National Accelerator Laboratory, Batavia, IL 60510, USA\\
$^{10}$Department of Physics, University of California, Santa Cruz, CA 95064 USA\\
$^{11}$Department of Physics and Astronomy, Pevensey Building, University of Sussex, Brighton, BN1 9QH, UK\\
$^{12}$Department of Physics, University of Arizona, Tucson, AZ 85721, USA\\
$^{13}$Department of Physics and Astronomy, University of Pennsylvania, Philadelphia, PA 19104, USA\\
$^{14}$Physics Department, University of Michigan, 450 Church Street, Ann Arbor, MI, 48109, USA\\
$^{15}$Cerro Tololo Inter-American Observatory, National Optical Astronomy Observatory, Casilla 603, La Serena, Chile\\
$^{16}$Department of Physics and Electronics, Rhodes University, PO Box 94, Grahamstown, 6140, South Africa\\
$^{17}$CNRS, UMR 7095, Institut d'Astrophysique de Paris, F-75014, Paris, France\\
$^{18}$Sorbonne Universit\'es, UPMC Univ Paris 06, UMR 7095, Institut d'Astrophysique de Paris, F-75014, Paris, France\\
$^{19}$Institute of Cosmology \& Gravitation, University of Portsmouth, Portsmouth, PO1 3FX, UK\\
$^{20}$Laborat\'orio Interinstitucional de e-Astronomia - LIneA, Rua Gal. Jos\'e Cristino 77, Rio de Janeiro, RJ - 20921-400, Brazil\\
$^{21}$Observat\'orio Nacional, Rua Gal. Jos\'e Cristino 77, Rio de Janeiro, RJ - 20921-400, Brazil\\
$^{22}$Department of Astronomy, University of Illinois, 1002 W. Green Street, Urbana, IL 61801, USA\\
$^{23}$National Center for Supercomputing Applications, 1205 West Clark St., Urbana, IL 61801, USA\\
$^{24}$Institut de Ci\`encies de l'Espai, IEEC-CSIC, Campus UAB, Carrer de Can Magrans, s/n,  08193 Bellaterra, Barcelona, Spain\\
$^{25}$School of Physics and Astronomy, University of Southampton,  Southampton, SO17 1BJ, UK\\
$^{26}$Excellence Cluster Universe, Boltzmannstr.\ 2, 85748 Garching, Germany\\
$^{27}$Faculty of Physics, Ludwig-Maximilians University, Scheinerstr. 1, 81679 Munich, Germany\\
$^{28}$Department of Astronomy, University of Michigan, Ann Arbor, MI 48109, USA\\
$^{29}$Kavli Institute for Cosmological Physics, University of Chicago, Chicago, IL 60637, USA\\
$^{30}$Department of Astronomy, University of California, Berkeley,  501 Campbell Hall, Berkeley, CA 94720, USA\\
$^{31}$Lawrence Berkeley National Laboratory, 1 Cyclotron Road, Berkeley, CA 94720, USA\\
$^{32}$Center for Cosmology and Astro-Particle Physics, The Ohio State University, Columbus, OH 43210, USA\\
$^{33}$Department of Physics, The Ohio State University, Columbus, OH 43210, USA\\
$^{34}$Australian Astronomical Observatory, North Ryde, NSW 2113, Australia\\
$^{35}$George P. and Cynthia Woods Mitchell Institute for Fundamental Physics and Astronomy, and Department of Physics and Astronomy, Texas A\&M University, College Station, TX 77843,  USA\\
$^{36}$Departamento de F\'{\i}sica Matem\'atica,  Instituto de F\'{\i}sica, Universidade de S\~ao Paulo,  CP 66318, CEP 05314-970, S\~ao Paulo, SP,  Brazil\\
$^{37}$Instituci\'o Catalana de Recerca i Estudis Avan\c{c}ats, E-08010 Barcelona, Spain\\
$^{38}$Jet Propulsion Laboratory, California Institute of Technology, 4800 Oak Grove Dr., Pasadena, CA 91109, USA\\
$^{39}$Centro de Investigaciones Energ\'eticas, Medioambientales y Tecnol\'ogicas (CIEMAT), Madrid, Spain\\
$^{40}$Argonne National Laboratory, 9700 South Cass Avenue, Lemont, IL 60439, USA

\label{lastpage}

\end{document}